\documentclass[]{interact}

\usepackage{epstopdf}% To incorporate .eps illustrations using PDFLaTeX, etc.

\usepackage{subcaption} % Provides the subfigure environment
\usepackage{placeins}
\usepackage[numbers,sort&compress]{natbib}% Citation support using natbib.sty
\bibpunct[, ]{[}{]}{,}{n}{,}{,}% Citation support using natbib.sty
\renewcommand\bibfont{\fontsize{10}{12}\selectfont}% Bibliography support using natbib.sty
\makeatletter% @ becomes a letter
\def\NAT@def@citea{\def\@citea{\NAT@separator}}% Suppress spaces between citations using natbib.sty
\makeatother% @ becomes a symbol again

\theoremstyle{plain}% Theorem-like structures provided by amsthm.sty

\theoremstyle{definition}

\theoremstyle{remark}

\usepackage[acronym]{glossaries}
\usepackage{tikz}
\usepackage{url}
\usetikzlibrary{shapes.geometric, arrows.meta, positioning}
\tikzstyle{block} = [rectangle, rounded corners=2pt, minimum width=3.5cm, minimum height=0.8cm, text centered, draw=black, fill=blue!10]
\tikzstyle{cblock} = [rectangle, rounded corners=2pt, minimum width=3.5cm, minimum height=0.8cm, text centered, draw=black, fill=orange!10]
\tikzstyle{iblock} = [rectangle, rounded corners=2pt, minimum width=3.5cm, minimum height=0.8cm, text centered, draw=black, fill=yellow!10]
\tikzstyle{oblock} = [rectangle, rounded corners=2pt, minimum width=3.5cm, minimum height=0.8cm, text centered, draw=black, fill=green!10]
\tikzstyle{qblock} = [rectangle, rounded corners=2pt, minimum width=3.5cm, minimum height=0.8cm, text centered, draw=black, fill=purple!15]
\tikzstyle{arrow} = [thick, -{Stealth}]
\usetikzlibrary{arrows.meta}
\usepackage{booktabs}
\usepackage{amssymb}
\usepackage{physics}
\usetikzlibrary{quantikz2}
\usepackage{amsmath}
\usepackage{subcaption}
\usepackage{hyperref}
\usepackage{cleveref}
\crefname{figure}{Figure}{Figures}
\Crefname{figure}{Figure}{Figures}
\crefname{table}{Table}{Tables}
\Crefname{table}{Table}{Tables}
\crefname{section}{Section}{Sections}
\Crefname{section}{Section}{Section}

\crefname{equation}{Equation}{Equations}
\Crefname{equation}{Equation}{Equations}

\newacronym{hqnns}{HQNNs}{Hybrid Quantum Neural Networks}
\newacronym{pqcs}{PQCs}{Parametrized Quantum Circuits}
\newacronym{cnn}{CNN}{Convolutional Neural Network}
\newacronym{qnn}{QNN}{Quantum Neural Network}
\newacronym{pca}{PCA}{Principal Component Analysis}
\newacronym{relu}{ReLU}{Rectified Linear Unit}
\newacronym{sgd}{SGD}{Stochastic Gradient Descent}

\usepackage{tikz}
\usepackage{scalerel}
\usetikzlibrary{svg.path}
\definecolor{orcidlogocol}{HTML}{A6CE39}
\tikzset{
  orcidlogo/.pic={
    \fill[orcidlogocol] svg{M256,128c0,70.7-57.3,128-128,128C57.3,256,0,198.7,0,128C0,57.3,57.3,0,128,0C198.7,0,256,57.3,256,128z};
    \fill[white] svg{M86.3,186.2H70.9V79.1h15.4v48.4V186.2z}
                 svg{M108.9,79.1h41.6c39.6,0,57,28.3,57,53.6c0,27.5-21.5,53.6-56.8,53.6h-41.8V79.1z M124.3,172.4h24.5c34.9,0,42.9-26.5,42.9-39.7c0-21.5-13.7-39.7-43.7-39.7h-23.7V172.4z}
                 svg{M88.7,56.8c0,5.5-4.5,10.1-10.1,10.1c-5.6,0-10.1-4.6-10.1-10.1c0-5.6,4.5-10.1,10.1-10.1C84.2,46.7,88.7,51.3,88.7,56.8z};
  }
}
\newcommand\orcidicon[1]{\href{https://orcid.org/#1}{\mbox{\scalerel*{
\begin{tikzpicture}[yscale=-1,transform shape]
\pic{orcidlogo};
\end{tikzpicture}
}{|}}}}

\begin{document}

%\articletype{ARTICLE TEMPLATE}% Specify the article type or omit as appropriate

\title{On the Importance of Fundamental Properties in Quantum-Classical Machine Learning Models}

\author{
\name{Silvie Ill\'{e}sov\'{a}\textsuperscript{a}\thanks{CONTACT Silvie Ill\'{e}sov\'{a}. Email: illesova.silvie.scholar@gmail.com} \orcidicon{0009-0002-5231-3714}
\and Tomasz Rybotycki\textsuperscript{b,c,d} \orcidicon{0000-0003-2493-0459}
\and Piotr Gawron\textsuperscript{c,d} \orcidicon{0000-0001-7476-9160}
\and Martin Beseda\textsuperscript{e} \orcidicon{0000-0001-5792-2872}}
\affil{\textsuperscript{a}IT4Innovations, VSB-Technical University of Ostrava, 17.~listopadu 2172/15, 708~00 Ostrava-Poruba, Czech Republic;
\textsuperscript{b}Systems Research Institute, Polish Academy of Sciences, Warszawa, Poland;
\textsuperscript{c}Nicolaus Copernicus Astronomical Center, Polish Academy of Sciences, Warszawa, Poland;
\textsuperscript{d}Center of Excellence in Artificial Intelligence, AGH University of Krakow, Cracow, Poland;
\textsuperscript{e}University of L'Aquila, Department of Information Engineering, Computer Science and Mathematics, L'Aquila, Italy}
}

\maketitle

\begin{abstract}
We present a systematic study of how quantum circuit design, specifically the depth of the variational ansatz and the choice of quantum feature mapping, affects the performance of hybrid quantum-classical neural networks on a causal classification task. The architecture combines a convolutional neural network for classical feature extraction with a parameterized quantum circuit acting as the quantum layer. We evaluate multiple ansatz depths and nine different feature maps. Results show that increasing the number of ansatz repetitions improves generalization and training stability, though benefits tend to plateau beyond a certain depth. The choice of feature mapping is even more critical: only encodings with multi-axis Pauli rotations enable successful learning, while simpler maps lead to underfitting or loss of class separability. Principal Component Analysis and silhouette scores reveal how data distributions evolve across network stages. These findings offer practical guidance for designing quantum circuits in hybrid models. All source codes and evaluation tools are publicly available.
\end{abstract}


\section{Introduction}

Quantum computing has rapidly gained attention as a transformative paradigm in computational science, with potential applications spanning physics~\cite{di2024quantum,joseph2023quantum, ciaramelletti2025detecting}, chemistry~\cite{illesova2025transformation, motta2022emerging, beseda2024state, cao2019quantum}, cryptography~\cite{pirandola2020advances, bennett1992experimental}, optimization\cite{illesova2025numerical, stilck2021limitations, shaydulin2019evaluating, novak2025optimization}, benchmarking~\cite{lewandowska2025benchmarking, bilek2025experimental, proctor2025benchmarking}, and machine learning~\cite{gupta2022quantum, schuld2015introduction, novak2025quantum}. By harnessing uniquely quantum phenomena such as superposition and entanglement, quantum computing also shows promise in enhancing classical machine learning models, leading to the development of hybrid approaches. 

\gls{hqnns}~\cite{zaman2024comparative, arthur2022hybrid, zeguendry2023quantum} have emerged as a promising approach to leverage the computational advantages of quantum computing within conventional machine learning workflows. These models integrate classical neural network components with \gls{pqcs}~\cite{du2020expressive, benedetti2019parameterized}, often referred to as quantum layers, enabling novel representations of data through the principles of superposition and entanglement. While the theoretical potential~\cite{cerezo2022challenges} of \gls{hqnns} is widely recognized, their practical success heavily depends on design choices related to the quantum circuit architecture and how classical data is encoded into quantum states~\cite{houssein2022machine}. 

Among the key design elements in \gls{hqnns} are the ansatz structures, specifically their depth, parameter count~\cite{lockwood2020reinforcement}, and the feature mapping strategy~\cite{kwon2024feature} that embeds classical inputs into quantum states. Deeper ansatz generally provide increased expressivity~\cite{nakaji2021expressibility} but may introduce challenges such as slower convergence, overfitting~\cite{kobayashi2022overfitting}, or barren plateaus~\cite{mcclean2018barren, qi2023barren} in the optimization landscape. Similarly, feature maps influence how effectively the quantum circuit can separate data classes; poorly chosen encodings can render the quantum model ineffective, even with optimal training~\cite{ranga2024quantum}.

Despite the growing number of hybrid models in the literature~\cite{metawei2020survey, de2022survey}, systematic studies that quantify the influence of ansatz depth and feature map choice on classification performance remain limited. In this work, we present a comprehensive evaluation of these two critical components using a series of controlled experiments on a hybrid neural network. We analyze the effects of varying ansatz repetitions and different feature maps on model accuracy, generalization ability, training stability, and data separability.

Our findings provide practical insights for the design of quantum circuits in machine learning. We show that while increasing ansatz depth can improve generalization and regularization, the marginal benefits diminish after a certain point. Moreover, only a subset of tested feature maps—specifically those incorporating multi-axis rotations—lead to successful learning. Dimensional collapse and poor clustering behavior are observed in models with suboptimal encodings, highlighting the importance of thoughtful feature map design.

By characterizing the interplay between circuit complexity and data encoding, this study contributes to a better understanding of trainability and performance in hybrid quantum-classical models. The presented results offer guidance for researchers developing next-generation quantum-enhanced learning systems. 

The paper\footnote{This work was presented at First Symposium on Physics and Chemistry for Unconventional Computing, in Krakow, Poland, 2025} is structured as follows. In the \cref{sec:methodology}, we describe the investigated model in depth, as well as all of the used metrics, which were chosen for the evaluation phase of our research. In the \cref{sec:results}, the focus is on the results themselves. Firstly, on the evaluation of the effect of ansatz depth on the model, and then on the in-depth analysis of feature mappings and their effects on the learning dynamics of the model. The conclusions are then presented in \cref{sec:conclusions}.

\section{Methodology}\label{sec:methodology}

This study investigates the effects of quantum circuit design choices, specifically, the depth of the parameterized ansatz and the choice of feature mapping, on the learning dynamics and performance of \gls{hqnns}. Our model architecture combines a \gls{cnn}~\cite{abdi1999neural, zou2009overview} with a \gls{qnn}~\cite{schuld2014quest} to perform multi-class classification. The classical component processes raw input data and transforms it into a lower-dimensional, more informative representation—called a feature vector—that captures relevant patterns or structures in the data. The quantum component applies \gls{pqcs} to the classical feature representation. First, a quantum feature map encodes the classical features into quantum states, embedding them into a high-dimensional Hilbert space~\cite{schuld2019quantum}. Then, trainable quantum gates act on these states, acting as a variational circuit whose parameters are optimized during training. This combination enables complex, non-linear transformations that leverage quantum phenomena such as superposition and entanglement, thereby enhancing the model's representational capacity beyond what classical networks alone can achieve\cite{abbas2021power}.

\subsection{\glsentryshort{cnn}}
The \gls{cnn} component of the hybrid model consists of three convolutional blocks, each designed to progressively extract more abstract spatial features from the input. These blocks increase the number of feature channels from 16 in the first layer to 32 and 64 in the subsequent layers, enabling the model to learn increasingly rich hierarchical representations.

Each block begins with a two-dimensional convolutional layer, which applies a set of learnable filters to the input tensor. The convolution operation uses a kernel of size $3 \times 3$ with a stride of one and a padding of one on all sides, preserving the spatial dimensions of the input. The choice of a $3 \times 3$ kernel with a stride of one and padding of one is motivated by the need to balance local feature extraction, spatial resolution preservation, and computational efficiency. A $3 \times 3$ kernel is widely used in convolutional neural networks~\cite{carlo1979spatial, jung2018extension} because it is small enough to capture fine-grained local patterns—such as edges or textures—while being large enough to aggregate information from neighboring pixels. Using a stride~\cite{zaniolo2020use} of one ensures that the convolution slides over the input one pixel at a time, allowing for dense sampling and detailed feature extraction. Padding~\cite{tang2019impact} of one on all sides compensates for the border shrinkage caused by the convolution, ensuring that the output has the same spatial dimensions as the input. This design preserves spatial information across layers, which is particularly important when working with small input sizes, such as $8 \times 8$ heatmaps, where even a slight reduction in resolution could significantly impact the model's capacity to retain structural patterns.
This kernel performs a local weighted sum over neighboring pixels, effectively computing a discrete convolution. Formally, for an input $X$ and kernel weights $W_k$, the output at position $(i,j)$ is given by
\begin{equation}
(X * W_k)(i, j) = \sum_{m=1}^{3} \sum_{n=1}^{3} X(i + m - 2, j + n - 2) \cdot W_k(m, n).
\end{equation}

Following the convolution, a \gls{relu}~\cite{banerjee2019empirical} activation function is applied. The \gls{relu} function introduces non-linearity by zeroing out all negative activations and leaving positive values unchanged, i.e.,
\begin{equation}
\text{ReLU}(x) = \max(0, x),
\end{equation}
which promotes sparse activations and improves gradient flow during training.

To progressively reduce the spatial resolution and concentrate the activations, each block includes a max pooling layer with a $2 \times 2$ window and a stride of 2. The choice of a $2 \times 2$ pooling window with a stride of 2 reflects a widely adopted design principle in convolutional architectures~\cite{ajit2020review}, where each pooling operation reduces the spatial resolution by a factor of two. This configuration strikes a balance between retaining essential spatial information and reducing the computational burden in deeper layers. It also ensures a gradual compression of the data~\cite{shadoul2025effect}, which helps the network generalize better by focusing on the most informative local features while discarding fine-grained noise. This operation partitions each channel of the feature map into non-overlapping $2 \times 2$ regions and replaces each region with its maximum value. Formally, let $X \in \mathbb{R}^{H \times W}$ denote a single-channel feature map. Then the pooled output $Y \in \mathbb{R}^{H/2 \times W/2}$ is computed as
\begin{equation}
Y(i, j) = \max_{\substack{0 \leq m < 2 \\ 0 \leq n < 2}} X(2i + m, 2j + n),
\end{equation}
for $0 \leq i < H/2$ and $0 \leq j < W/2$.

This downsampling reduces the spatial dimensions by a factor of two in each direction and retains the most salient feature within each region, contributing to translational invariance and reduced computational complexity in subsequent layers.

Finally, dropout with a probability of $0.5$ is applied after each pooling layer to prevent overfitting. This value is chosen based on previous analysis~\cite{park2017analysis, wu2015max}. Dropout randomly deactivates a subset of the neurons during training according to a Bernoulli distribution, effectively zeroing out each activation $x_i$ with probability $p = 0.5$, and scaling the remaining activations by $1/(1-p)$ to preserve the expected value
\begin{equation}
    \tilde{x}_i = \frac{m_i \cdot x_i}{1 - p}, \quad m_i \sim \text{Bernoulli}(1 - p).
\end{equation}

This regularization technique encourages the model to learn redundant representations and reduces sensitivity to individual neurons~\cite{yeung2010sensitivity, park2017analysis}.

After passing through all three convolutional blocks, the feature maps are reshaped into one-dimensional vectors using a flattening operation that preserves the batch dimension. For input images of size $8 \times 8$, each pooling layer reduces the spatial resolution by a factor of two, resulting in feature maps of size $1 \times 1$ at the final layer. Since the final block has 64 channels, this yields a flattened vector of length 64 per sample.

To interface with the quantum component of the model, this 64-dimensional representation is projected down to a lower-dimensional space via a fully connected linear layer. Specifically, a linear transformation $\mathbb{R}^{64} \rightarrow \mathbb{R}^{n_q}$ is applied, where $n_q$ denotes the number of qubits used in the quantum layer. 
The meaning of $n_q$ remains the same throughout the whole paper, always signifying the number of qubits in the inserted quantum layer. And for all experiments, the number of qubits has been set to $n_q = 3$. This dimensionality reduction ensures that each input sample is mapped to a vector of length $n_q$, which is compatible with the input requirements of the quantum feature map and ansatz, enabling effective state preparation in the quantum circuit. The whole process is shown in \cref{tab:cnn-dim-reduction}.

\begin{table}[ht]
\centering
\caption{Dimensionality reduction through the CNN layers for input size $1 \times 8 \times 8$}
\begin{tabular}{|l|c|c|}
\hline
\textbf{Layer} & \textbf{Operation} & \textbf{Output Shape} \\
\hline
Input & -- & $1 \times 8 \times 8$ \\
\hline
Conv1 (kernel $3\times3$, padding 1) & Convolution & $16 \times 8 \times 8$ \\
\hline
Pool1 (kernel $2\times2$, stride 2) & Max pooling & $16 \times 4 \times 4$ \\
\hline
Conv2 (kernel $3\times3$, padding 1) & Convolution & $32 \times 4 \times 4$ \\
\hline
Pool2 (kernel $2\times2$, stride 2) & Max pooling & $32 \times 2 \times 2$ \\
\hline
Conv3 (kernel $3\times3$, padding 1) & Convolution & $64 \times 2 \times 2$ \\
\hline
Pool3 (kernel $2\times2$, stride 2) & Max pooling & $64 \times 1 \times 1$ \\
\hline
Flatten & Reshape & $64$ \\
\hline
Linear & Fully connected $\mathbb{R}^{64} \rightarrow \mathbb{R}^{n_q}$ & $n_q$ \\
\hline
\end{tabular}
\label{tab:cnn-dim-reduction}
\end{table}

\begin{figure}[ht!]
\centering
\begin{tikzpicture}[node distance=0.3cm]

% Classical blocks
\node[iblock] (input) {Step 1: Input Image};
\node[cblock, below=of input] (cnn) {\shortstack{Step 2: CNN Feature Extractor\\(Conv + ReLU + Pool + Dropout)}};
\node[block, below=of cnn] (flatten) {Step 3: Flatten};

% Linear projection
\node[block, below=of flatten] (reduce) {\shortstack{Step 4: Dimensionality Reduction\\(Linear Layer)}};

% Quantum block
\node[qblock, below=of reduce] (qnn) {\shortstack{Step 5: Quantum Neural Network\\(Feature Map + Ansatz)}};

% Output
\node[block, below=of qnn] (classifier) {\shortstack{Step 6: Final Classifier\\(Linear Layer)}};
\node[oblock, below=of classifier] (output) {Step 7: Output Class Scores};

% Arrows
\draw[arrow] (input) -- (cnn);
\draw[arrow] (cnn) -- (flatten);
\draw[arrow] (flatten) -- (reduce);
\draw[arrow] (reduce) -- (qnn);
\draw[arrow] (qnn) -- (classifier);
\draw[arrow] (classifier) -- (output);

\end{tikzpicture}
\caption{Step-wise architecture of the hybrid quantum-classical model. Classical processing (blue) extracts and projects features, quantum processing (purple) performs parameterized evolution, and the final classifier outputs class probabilities.}
\label{fig:hybridqnn}
\end{figure}

\subsection{QNN}
The quantum layer is constructed using \textit{Qiskit Machine Learning}\footnote{\url{https://qiskit-community.github.io/qiskit-machine-learning/index.html}}~\cite{sahin2025qiskit} and comprises two main components. A data encoding circuit (feature map) and a parameterized ansatz. Feature maps encode classical input vectors $x \in \mathbb{R}^{n_q}$ into quantum states by applying data-dependent unitary operations to an initial reference state, typically $\ket{0}^{\otimes n_q}$.

One of the simplest encoding strategies is the use of $Z$-rotation gates of the form
\begin{equation}
R_z(\theta) = \exp\left(-i \frac{\theta}{2} Z\right) =
\begin{bmatrix}
e^{-i\theta/2} & 0 \\
0 & e^{i\theta/2}
\end{bmatrix},
\label{eq:PauliZ}
\end{equation}
where $\theta$ is a function of a classical input component, typically $\theta = x_i$ for the $i$-th qubit. This gate applies a rotation around the $Z$ axis of the Bloch sphere and embeds the input into the quantum phase.

More expressive encodings include entangling operations such as the two-qubit \texttt{ZZ} gate, which is used in circuits like \texttt{ZZFeatureMap}. The \texttt{ZZ} effect corresponds to the unitary operator
\begin{equation}
\text{ZZ}(\theta) = \exp\left(-i \frac{\theta}{2} Z \otimes Z\right) =
\begin{bmatrix}
e^{-i\theta/2} & 0 & 0 & 0 \\
0 & e^{i\theta/2} & 0 & 0 \\
0 & 0 & e^{i\theta/2} & 0 \\
0 & 0 & 0 & e^{-i\theta/2}
\end{bmatrix}.
\label{eq:zz}
\end{equation}
This gate entangles pairs of qubits via the product of Pauli-$Z$ operators, enabling input-dependent correlations to be encoded across qubits.

Additionally, feature maps may include single-qubit $R_x$ and $R_y$ rotations, whose matrix forms are given respectively by
\begin{equation}
R_x(\theta) = \exp\left(-i \frac{\theta}{2} X\right) =
\begin{bmatrix}
\cos(\theta/2) & -i\sin(\theta/2) \\
-i\sin(\theta/2) & \cos(\theta/2)
\end{bmatrix},
\label{eq:PauliX}
\end{equation}
\begin{equation}
R_y(\theta) = \exp\left(-i \frac{\theta}{2} Y\right) =
\begin{bmatrix}
\cos(\theta/2) & -\sin(\theta/2) \\
\sin(\theta/2) & \cos(\theta/2)
\end{bmatrix}.
\label{eq:PauliY}
\end{equation}
These gates provide additional expressive power in the embedding by varying the rotation axis used for encoding.

Together, these operations form the data-dependent unitary $U_{\text{feat}}(x)$ that maps classical inputs to quantum states
\begin{equation}
\ket{\psi(x)} = U_{\text{feat}}(x) \ket{0}^{\otimes n_q},
\end{equation}
preparing an input-specific quantum state for subsequent processing by the variational circuit.

The ansatz is implemented using a \texttt{TwoLocal}\footnote{\url{https://docs.quantum.ibm.com/api/qiskit/qiskit.circuit.library.TwoLocal}} circuit with linear entanglement and configurable depth. The \texttt{TwoLocal} ansatz is a parameterized variational circuit that alternates between layers of single-qubit rotation gates and entangling two-qubit gates. 

Formally, for a system of $n_q$ qubits and depth $d$, the circuit has the form
\begin{equation}
U_{\text{ansatz}}(\boldsymbol{\theta}) = \left( \mathcal{E} \cdot \mathcal{R}(\boldsymbol{\theta}_l) \right)^d,
\end{equation}
% \si{@Tomek, this, I believe, is not data reuploading, as in this part I am talking about the ansatz, and increasing the number of parameters in it by repetition of ansatz blocks. In my understanding, data reuploading would be a case where we have several repetitions of the feature map, enforcing the data that we encode. Feel free to correct me if I am mistaken in my understanding :)}
where $\mathcal{R}(\boldsymbol{\theta}_l)$ is a layer of parameterized single-qubit rotations applied to each qubit, and $\mathcal{E}$ is an entangling layer defined by a fixed topology. If we choose a setting, shown in \cref{fig:twolocal-circuit}, where the rotation layer applies $R_y$ and $R_z$ gates
\begin{equation}
\mathcal{R}(\theta_i, \phi_i) = R_y(\theta_i) R_z(\phi_i), \quad \text{for each qubit } i,
\end{equation}
where $ R_y(\theta_i), R_z(\phi_i)$ are Pauli rotation gates from \cref{eq:PauliZ,eq:PauliY}.

The entanglement layer $\mathcal{E}$ is composed of \texttt{CX} (controlled-NOT) gates arranged in a \emph{linear} configuration, each qubit is entangled with its immediate neighbor, i.e., qubit $i$ is entangled with qubit $i+1$ for $i = 0, \ldots, n_q - 2$.

This architecture yields a compact yet expressive variational circuit with $2dn_q$ trainable parameters when two rotation gates are used per qubit per layer. The overall unitary can be written as
\begin{equation}
U_{\text{ansatz}}(\boldsymbol{\theta}) = \prod_{l=1}^{d} \left( \mathcal{E} \cdot \mathcal{R}(\boldsymbol{\theta}_l) \right),
\end{equation}
where each layer of entanglement and rotations introduces a learnable structure into the quantum state.

\begin{figure}[ht!]
\centering
\begin{quantikz}[row sep=0.4cm, column sep=0.4cm]
\lstick{$\ket{0}$} & \gate{R_y(\theta_1)} & \gate{R_z(\phi_1)} & \ctrl{1} & \qw       & \qw \\
\lstick{$\ket{0}$} & \gate{R_y(\theta_2)} & \gate{R_z(\phi_2)} & \targ{}  & \ctrl{1}  & \qw \\
\lstick{$\ket{0}$} & \gate{R_y(\theta_3)} & \gate{R_z(\phi_3)} & \qw      & \targ{}   & \qw
\end{quantikz}
\caption{One repetition (depth = 1) of a \texttt{TwoLocal} ansatz with $R_y$ and $R_z$ rotations and linear CNOT entanglement for $n_q = 3$ qubits.}
\label{fig:twolocal-circuit}
\end{figure}

As the learning task does not assume a fixed physical interpretation or quantum advantage rooted in domain-specific structure, we adopt a hardware-efficient ansatz\cite{leone2024practical, wang2024entanglement} to ensure a favorable balance between expressivity and trainability. Such circuits are designed to be compatible with near-term quantum devices and minimize circuit depth by using shallow layers of parameterized single-qubit rotations and entangling gates arranged in native hardware topologies.

The quantum circuit is integrated into the \texttt{PyTorch} framework\cite{NEURIPS2019_9015} via the \texttt{TorchConnector} interface provided by Qiskit Machine Learning. This integration enables end-to-end training using classical optimization techniques, including stochastic gradient descent and its variants.

To support gradient-based learning, we enable the computation of input gradients concerning the quantum model in the \texttt{EstimatorQNN}. This ensures that the quantum layer supports differentiation not only with respect to its trainable parameters, but also with respect to the classical inputs $x$ that are embedded via the feature map. Formally, given a quantum model output $f(x; \theta)$, this allows the computation of gradients of the form $\nabla_x f(x; \theta)$ in addition to $\nabla_\theta f(x; \theta)$.

These gradients are computed using the parameter-shift rule\cite{wierichs2022general,markovich2024parameter}, a method compatible with variational quantum circuits that enables exact gradient computation by evaluating the circuit at shifted parameter values
\begin{equation}
\frac{\partial f}{\partial \theta_i} = \frac{1}{2} \left[ f(\theta_i + \frac{\pi}{2}) - f(\theta_i - \frac{\pi}{2}) \right].
\end{equation}
This approach allows the quantum layer to be trained within the automatic differentiation graph of PyTorch, ensuring compatibility with modern deep learning workflows.

\subsection{Dataset}
The dataset used for training and evaluation originates from the Kaggle “Cause-Effect Pairs” competition and is publicly available online\footnote{\url{https://www.kaggle.com/c/cause-effect-pairs}}. Each input sample is a bivariate dataset containing two random variables $ X \in \mathbb{R}^n $ and $ Y \in \mathbb{R}^n $, where the goal is to infer the causal direction between them. From each variable pair, we compute a joint histogram over a fixed grid to produce an image-based representation of their statistical relationship. The dataset consisted of 1024 samples for the training phase and 1024 for the validation.

Concretely, for each pair $ (X, Y) $, we construct an $8 \times 8$ normalized heatmap $ H \in [0,1]^{8 \times 8} $ where each entry $ H_{i,j} $ corresponds to the (normalized) count of samples falling into the $i$-th and $j$-th bins of an equally spaced discretization of the $X$ and $Y$ axes, respectively. This produces a fixed-size, image-like matrix encoding the joint distribution between variables. These heatmaps are treated as grayscale images and serve as the input to the hybrid model.

Each heatmap is labeled with one of three mutually exclusive classes: positive causality if $X \rightarrow Y$, negative causality if $Y \rightarrow X$, or no causality if the relationship is judged to be acausal or confounded. These categorical labels serve as supervision targets during training.

Before training, all heatmap values are normalized to the range $[0, 1]$ by dividing each entry by the maximum count in the original (unnormalized) histogram
\begin{equation}
   H^\text{norm}_{i,j} = \frac{H_{i,j}}{\max_{i,j} H_{i,j}}, 
\end{equation}
ensuring consistency in scale across samples and improving numerical stability during training.

These normalized heatmaps serve as fixed-length representations of variable interactions. By training on this format, the model is encouraged to learn statistical cues indicative of \emph{causal structure}—defined here as the directional dependence between observed variables—in an end-to-end manner. We evaluate how effectively different quantum circuit designs (i.e., alternative feature maps and ansatz structures) can capture these causal relationships in a hybrid learning setting.

\subsection{Training}
All models are trained using the cross-entropy loss function for multi-class classification. Cross-entropy loss is used for multi-class classification because it measures the dissimilarity between the predicted probability distribution and the true label distribution~\cite{mannor2005cross}, which is typically one-hot encoded. It penalizes incorrect predictions more strongly when the model is confident but wrong, encouraging well-calibrated probability estimates~\cite{shore2003properties, golik2013cross}. This makes it particularly suitable for classification tasks, as it directly optimizes the likelihood of the correct class, leading to faster and more stable convergence during training. Given a predicted probability distribution $\hat{y} \in \mathbb{R}^{C}$ over $C$ classes and a one-hot encoded ground truth label $y \in \{0,1\}^C$, the loss is computed as
\begin{equation}
\mathcal{L}_{\text{CE}}(y, \hat{y}) = -\sum_{c=1}^{C} y_c \log(\hat{y}_c),
\end{equation}
where $\hat{y}_c$ denotes the softmax probability assigned to class $c$.

Optimization is carried out using \gls{sgd} with Nesterov momentum~\cite{amari1993backpropagation,qu2019accelerated}. The parameter update rule at iteration $t$ is given by
\begin{align}
v_{t+1} &= \mu v_t - \eta \nabla_\theta \mathcal{L}(\theta_t + \mu v_t), \\
\theta_{t+1} &= \theta_t + v_{t+1},
\end{align}
where $\eta$ is the learning rate, $\mu$ is the momentum coefficient (set to 0.9), and $v_t$ is the accumulated velocity vector. Nesterov momentum provides a lookahead effect by computing the gradient at the anticipated next position, leading to faster and more stable convergence compared to vanilla momentum.

A small $L_2$ regularization term (weight decay) with a coefficient of $10^{-6}$ is added to the loss to prevent overfitting. This encourages the model to prefer smaller weight magnitudes by modifying the loss function as
\begin{equation}
\mathcal{L}_{\text{total}} = \mathcal{L}_{\text{CE}} + \lambda \|\theta\|_2^2,
\end{equation}
where $\lambda = 1 \times 10^{-6}$. This regularization technique penalizes overly complex models and helps improve generalization.

The learning rate is initialized to $\eta = 0.01$, which provides a balance between fast convergence and stable updates~\cite{iyer2023maximal}, especially in the early epochs when gradients are large. The learning rate $\eta = 0.01$ is selected to ensure sufficiently large updates for rapid learning while avoiding divergence. The batch size is set to 64, a common choice that balances convergence stability with computational efficiency on modern GPUs~\cite{radiuk2017impact,masters2018revisiting}. These hyperparameters were selected based on empirical conventions in convolutional and hybrid quantum-classical models and could be further tuned in future work.

Model convergence is monitored using validation accuracy and early stopping. Specifically, validation accuracy is computed after each epoch, and training is terminated if no improvement is observed for 100 consecutive epochs (patience = 100). Saying that if there is no improvement in any subsequent 100 epochs, we terminate the calculation. This prevents overfitting to the training set while ensuring sufficient training time for convergence.

\subsection{Performance metrics}
Performance metrics, including training and validation accuracy, are logged at the end of every epoch. This epoch-level logging allows for post hoc analysis of convergence behavior, learning dynamics, and generalization performance.

A comprehensive evaluation framework is employed to characterize the behavior and learning dynamics of each model. Standard metrics include training accuracy and validation accuracy~\cite{bustillo2022improving}, defined  as
\begin{equation}
\text{A} = \frac{1}{N} \sum_{i=1}^N \mathbb{I}[\hat{y}_i = y_i],
\end{equation}
where $N$ is the number of samples, $y_i$ is the ground truth label, $\hat{y}_i$ is the predicted label, and $\mathbb{I}[\cdot]$ is the indicator function.

To gain deeper insight into generalization and learning stability, several diagnostic indicators are also computed. One such metric is the \emph{generalization gap}~\cite{wang1994theory}, defined as the absolute difference between training and validation accuracy
\begin{equation}
\Delta_{\text{gen}} = \left| \text{A}_{\text{train}} - \text{A}_{\text{val}} \right|.
\label{eq:gen_gap}
\end{equation}
A larger generalization gap typically indicates overfitting~\cite{ying2019overview}, where the model fits the training data well but fails to generalize to unseen samples.

The \emph{early learning slope} is calculated as the change in validation accuracy over the first $k$ epochs
\begin{equation}
\text{S}_{\text{e}} = \frac{\text{A}_{\text{val}}^{(k)} - \text{A}_{\text{val}}^{(1)}}{k - 1},
\label{eq:slope}
\end{equation}
where $\text{A}_{\text{val}}^{(i)}$ denotes the validation accuracy at epoch $i$. This slope provides a coarse estimate of the model’s initial learning speed and responsiveness to the optimization signal.

The \emph{overfitting drop}~\cite{dietterich1995overfitting} quantifies the decline in validation accuracy after the model reaches its peak performance
\begin{equation}
\text{D}_{\text{overfit}} = \text{A}_{\text{val}}^{\text{peak}} - \text{A}_{\text{val}}^{\text{final}},
\label{eq:drop_overfit}
\end{equation}
where $\text{A}_{\text{val}}^{\text{peak}} = \max_{i} \text{A}_{\text{val}}^{(i)}$ is the highest observed validation accuracy across all epochs, and $\text{A}_{\text{val}}^{\text{final}}$ is the accuracy at the final epoch. A large drop may signal instability or overfitting in the later stages of training.

Together, these metrics and diagnostics offer a view of model performance, helping to identify not only which models perform best but also how they learn and generalize over time.

Additional indicators, such as fluctuation metrics, defined by local standard deviation of accuracy, and the stability ratio, which compares fluctuations in training and validation performance, are also computed. The  standard deviation $\sigma_{\Delta}$ and mean $\mu_{\Delta}$ of these differences were used to quantify fluctuation magnitude and are defined as
\begin{equation}
    \mu_{\Delta} = \frac{1}{T - 1} \sum_{t=1}^{T-1} |\Delta_t|,
    \label{eq:mu}
\end{equation}
\begin{equation}
    \sigma_{\Delta} = \sqrt{ \frac{1}{T - 1} \sum_{t=1}^{T-1} (|\Delta_t| - \mu_{\Delta})^2 },
    \label{eq:sigma}
\end{equation}
where $\Delta_t = a_{t+1} - a_t$ is the change in accuracy between epochs.
The stability ratio is defined as
\begin{equation}
R_{\text{stability}} = \frac{\mu^{\text{val}}_{\Delta}}{\mu^{\text{train}}_{\Delta}}.
\label{eq:stabi}
\end{equation}

Beyond accuracy-based metrics, data separability is assessed using an unsupervised clustering tool, as there is a need to examine how different parts of our model are changing the data structure. \gls{pca} is applied at several network stages to visualize how the data distribution evolves~\cite{abdi2010principal,greenacre2022principal}. The Silhouette score~\cite{shahapure2020cluster} is computed after each \gls{pca} transformation to quantify cluster compactness and class separability. To calculate the Silhouette score, let us first define the Silhouette coefficient
\begin{equation}
    s(i) = \frac{b(i) - a(i)}{\max \{ a(i), b(i) \}},
\end{equation}
where $a(i)$ is the mean distance between point $i$ and all other points in the same cluster, we call it the Compactness of the cluster and define it later. The Separability $b(i)$ is the mean distance between point $i$ and all points in the nearest neighboring cluster, the one to which point $i$ does not belong and has the lowest average distance. 

Then the Silhouette score $S$ for the whole dataset is obtained by 
\begin{equation}
    S = \frac{1}{n} \sum_{i=1}^{n} s(i),
\end{equation}
where $n$ is the number of samples in the whole dataset. By compactness of the cluster, we mean the cohesion of the samples assigned to the cluster, and it can be defined as
\begin{equation}
    a(i) = \frac{1}{|C_i| - 1} \sum_{\substack{j \in C_i \\ j \ne i}} d(i, j),
\end{equation}
where $C_i$ is the cluster into which point $i$ belong to, $|C_i|$ is the size, number of points, in cluster $C_i$ and $d(i,j)$ is the Euclidean distance between points $i$ and $j$.

As for Separability, we define it as
\begin{equation}
    b(i) = \min_{k \ne C_i} \left( \frac{1}{|C_k|} \sum_{j \in C_k} d(i, j) \right),
\end{equation}
here $C_k$ is the cluster into which point $i$ does not belong to.

To isolate the impact of individual architectural elements, such as the classical part, the feature map, and the quantum layer, experiments are conducted under controlled conditions. Multiple quantum ansatz depths are tested while keeping all other components unchanged. Analogously, different feature maps are compared under a fixed model configuration. Each experimental configuration is independently trained and evaluated, and all results, the accuracies in every epoch, the model itself, and after the last epoch, the intermediate results for \gls{pca} at three different stages of the model, after the classical part, after the feature mapping, and at the output of the quantum layer. They are saved for later comparison and visualization. This experimental design allows a systematic analysis of how architectural choices in the quantum layer affect the overall behavior of the hybrid model.

 The complete source code is publicly available for reproducibility via Zenodo~\cite{Illesova2025QuantumLayers} and on GitLab\footnote{https://gitlab.com/illesova.silvie.scholar/leveraging-quantum-layers-in-classical-neural-networks}. The repository includes all scripts used for training, evaluation, visualization, and figure generation.

\section{Results}\label{sec:results}
The result section is divided into two parts, the first of which is focused on the effects of ansatz depth on the learning process. There, we evaluate three models that differ only in the number of blocks in the ansatz. The focus of this part is on the learning stability and the evaluation of potential overfitting of the models. In the second part, we evaluate different feature mappings, having a deeper look at how they impact the separability of data clusters and their impact on the overall learning dynamics.

\subsection{Effect of Ansatz Depth on Learning Dynamics}

To evaluate the influence of quantum circuit depth on hybrid model performance, we varied the number of ansatz repetitions from one to three while keeping all other architectural and training parameters fixed. Each model was trained from scratch and evaluated using the accuracy of the predictions, generalization gap, \cref{eq:gen_gap}, and training stability, \cref{eq:stabi}.

Final accuracy results are presented in \cref{tab:final_accuracy}. Increasing the number of ansatz repetitions led to consistent improvements in both training and validation accuracy until the last case of five repetitions, where the accuracy fell again, which points to the fact that there is an optimal point for this parameter. Raising the number of ansatz repetitions above said point does not yield any advantage, as we get both lower training and validation accuracy than in the case of two- or three-repetitions. The most significant gain was observed when increasing the depth from one to two repetitions. Although the performance gain from two to three repetitions was smaller, the model with three repetitions achieved the highest validation accuracy overall, suggesting improved generalization with somewhat deeper circuits. 

\begin{table}[ht!]
\centering
\caption{Final Accuracy Metrics}
\label{tab:final_accuracy}
\begin{tabular}{lcc}
\toprule
\textbf{Repetitions} & \textbf{Training Accuracy} & \textbf{Validation Accuracy} \\
\midrule
1 Repetition & 0.8467 & 0.8721 \\
2 Repetitions & 0.9121 & 0.8955 \\
3 Repetitions & 0.9004 & 0.9111 \\
5 Repetitions & 0.8447 & 0.8857\\
\bottomrule
\end{tabular}
\end{table}

\begin{figure}[ht!]
\centering
\includegraphics[width=0.95\textwidth]{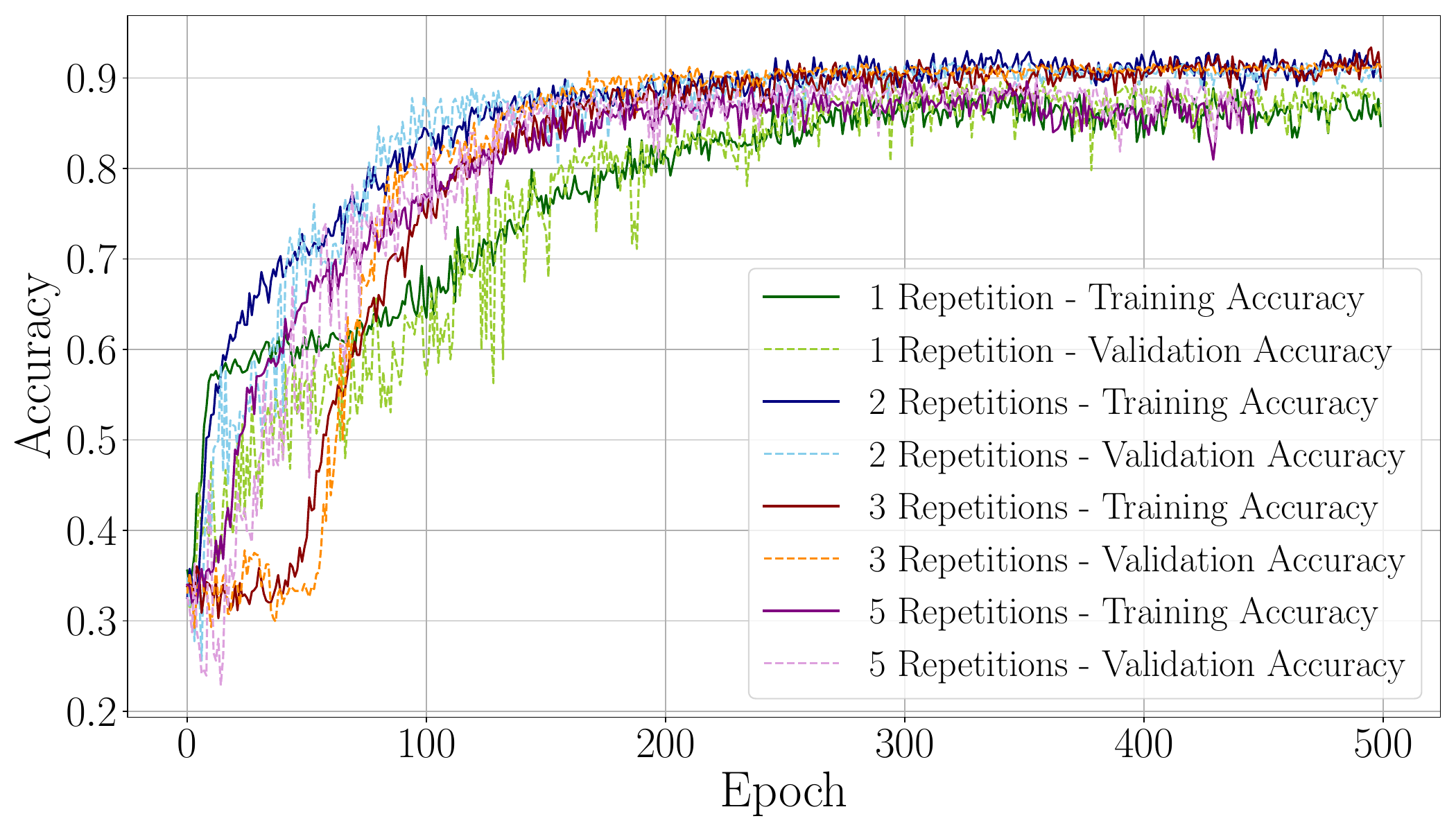}
\caption{Training and validation accuracy curves for models with 1, 2, 3, and 5 ansatz repetitions.}
\label{fig:train_val_accuracy}
\end{figure}

\cref{fig:train_val_accuracy} shows the training and validation accuracy curves for all three models over 500 training epochs. While all the models exhibit increasing accuracy over time, distinct training dynamics emerge. The model with one repetition exhibits a steep initial increase in both training and validation accuracy, but plateaus early and shows pronounced fluctuations, particularly in the accuracy considering the validation data set. The two-repetition model demonstrates smoother convergence and improved stability. The three-repetition model displays an initial plateau in early epochs but then surpasses the other models with both higher accuracy and lower fluctuation, indicating that deeper circuits stabilize training despite a slower start. We chose to examine up to three repetitions as we observed that both 2 and 3 repetitions are converging to a similar level of accuracy, and an increase in the number of parameters that the additional repetition would add would add additional complexity, while providing diminishing returns. To further analyze this trend, we additionally evaluated an ansatz scenario with five repetitions, which achieves higher accuracy than the single-repetition model but does not improve upon the two- and three-repetition cases, suggesting that increased depth alone does not guarantee better performance and further complicates optimization, as it increases the number of dimensions in the search space.

To quantify the generalization property, the absolute generalization gap, \cref{eq:gen_gap}, is defined as the difference between training and validation accuracy.

% \begin{equation}
% \text{G}_{\text{gap}} = |A_{\text{train}} - A_{\text{val}}|.
% \end{equation}

\begin{figure}[ht!]
\centering
\includegraphics[width=0.9\textwidth]{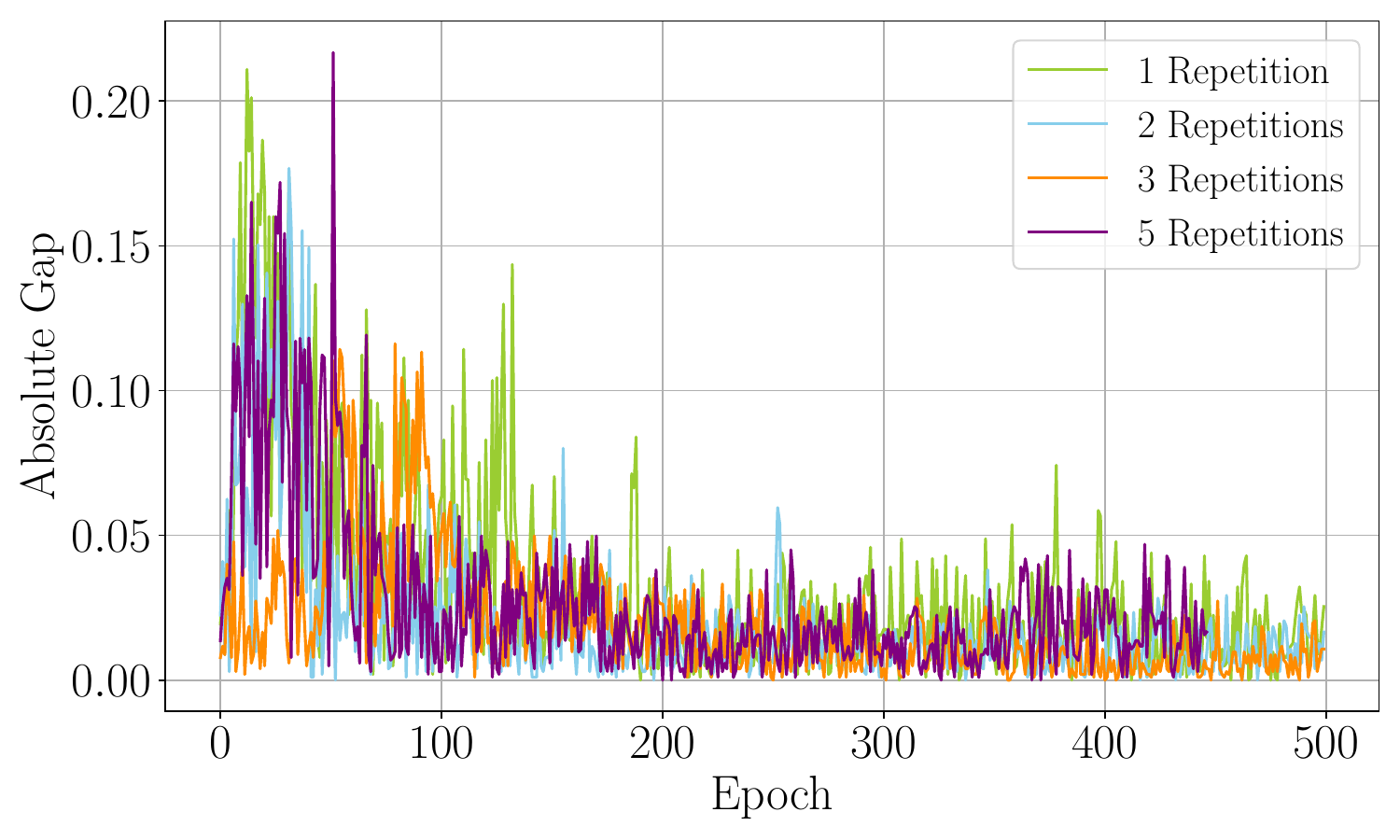}
\caption{Absolute generalization gap during training across different ansatz depths.}
\label{fig:generalization_gap}
\end{figure}

\cref{fig:generalization_gap} illustrates this metric over time, and the final and mean values are shown in \cref{tab:gap_metrics}. The model with one repetition exhibits larger and more volatile gaps, especially in early training. The models with two and three repetitions display reduced fluctuation and better alignment between training and validation accuracy, reflecting improved generalization. However, the model with 5 repetitions shows worse results, in line with the observations made before.

\begin{table}[ht!]
\centering
\caption{Generalization Gap Metrics}
\label{tab:gap_metrics}
\begin{tabular}{lcc}
\toprule
\textbf{Repetitions} & \textbf{Final Gap} & \textbf{Mean Gap} \\
\midrule
1 Repetition & 0.0254 & 0.0332 \\
2 Repetitions & 0.0166 & 0.0200 \\
3 Repetitions & 0.0107 & 0.0195 \\
5 Repetitions & 0.0410 & 0.0273 \\
\bottomrule
\end{tabular}
\end{table}

To assess learning efficiency, we considered the epoch at which the model first achieved 90\% validation accuracy, the early learning slope over the first five epochs, and the overfitting drop. These were computed using \cref{eq:drop_overfit,eq:slope} with the addition of 
\begin{equation}
    \text{E}_{90\%} = \min \{ t \mid A_{\text{val}}(t) \geq 0.9 \}.
\end{equation}
% \begin{align}
%     \text{E}_{90\%}& = \min \{ t \mid A_{\text{val}}(t) \geq 0.9 \},\\
%     \text{D}_{\text{early}} &= \frac{A_{\text{val}}(t=N) - A_{\text{val}}(t=0)}{N}, \quad N=5,\\
%     \text{Drop}_{\text{overfit}} &= A_{\text{peak}} - A_{\text{final}}.
% \end{align}

The results of these metrics are described in \cref{tab:learning_efficiency}.

\begin{table}[ht!]
\centering
\caption{Learning Efficiency and Overfitting Metrics}
\label{tab:learning_efficiency}
\begin{tabular}{lccc}
\toprule
\textbf{Repetitions} & \textbf{Epoch@90\%} & \textbf{Early Slope} & \textbf{Overfitting Drop} \\
\midrule
1 Repetition & Not reached & 0.0117 & 0.0244 \\
2 Repetitions & 195 & -0.0068 & 0.0234 \\
3 Repetitions & 168 & -0.0051 & 0.0068 \\
5 Repetitions & 355 & -0.0025 & 0.0147\\
\bottomrule
\end{tabular}
\end{table}

To further characterize training stability, we measured local fluctuations in training accuracy by computing first-order differences between adjacent epochs. The standard deviation $\sigma_{\Delta}$ and mean $\mu_{\Delta}$ are defined in \cref{eq:mu,eq:sigma}. The results for individual runs are shown in \cref{tab:train_fluct}.

\begin{table}[ht!]
\centering
\caption{Training Accuracy Fluctuation Metrics}
\label{tab:train_fluct}
\begin{tabular}{lcc}
\toprule
\textbf{Repetitions} & \textbf{Std. Dev.} & \textbf{Mean Abs. Diff.} \\
\midrule
1 Repetition & 0.0106 & 0.0122 \\
2 Repetitions & 0.0082 & 0.0103 \\
3 Repetitions & 0.0083 & 0.0105 \\
5 Repetitions & 0.0106 & 0.0130\\
\bottomrule
\end{tabular}
\end{table}

Validation accuracy fluctuations and the derived stability ratio are shown in \cref{tab:val_fluct}. The stability ratio is given by \cref{eq:stabi} and a value less than one indicates more stable performance on the validation set than on the training set, which is a desirable property for generalization, as a more stable learning process is less prone to overfitting. While the stability ratio does not directly tell us how effective the training is, it tells us that we are obtaining results with the same degree of accuracy for both the training and validation data.

\begin{table}[ht!]
\centering
\caption{Validation Accuracy Fluctuation and Stability Ratio}
\label{tab:val_fluct}
\begin{tabular}{lccc}
\toprule
\textbf{Number of Repetitions} & \textbf{Val Std. Dev.} & \textbf{Mean Abs. Diff.} & \textbf{Stability Ratio} \\
\midrule
1  & 0.0288 & 0.0265 & 2.1791 \\
2  & 0.0207 & 0.0158 & 1.5353 \\
3  & 0.0131 & 0.0088 & 0.8335 \\
5 & 0.0259 & 0.0209 & 1.6068\\
\bottomrule
\end{tabular}
\end{table}

In summary, increasing the depth of the ansatz yields improvements in validation accuracy, generalization gap, and training stability. While deeper circuits introduce a short delay in early learning, they enhance consistency and reduce overfitting. These findings imply that deeper quantum circuits serve as implicit regularizers, improving not only model outcomes but also training dynamics.

\subsection{Impact of Feature Mapping on Hybrid Model Performance}
The choice of feature mapping plays a pivotal role in determining the performance of hybrid quantum-classical neural networks.

\begin{figure}
    \centering
    \includegraphics[width=0.5\linewidth]{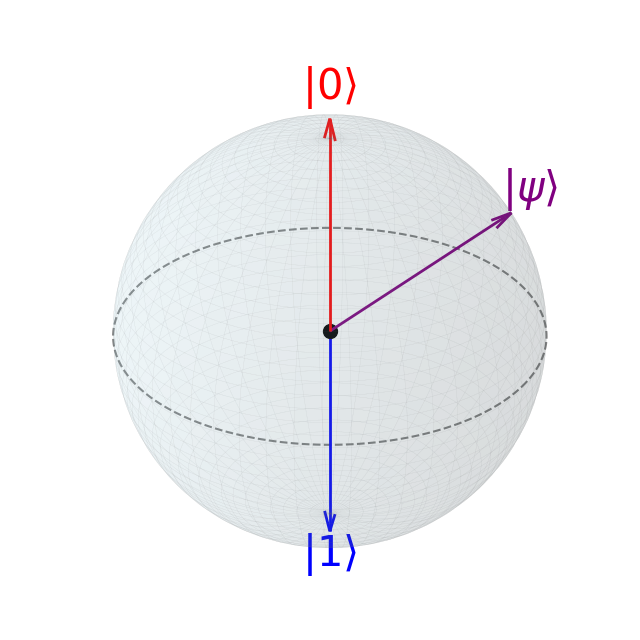}
    \caption{Bloch sphere, which is used to visualize the single qubit state}
    \label{fig:bloch}
\end{figure}

Several different feature mappings have been investigated. Throughout this manuscript a following convention is used to distinguish between nine. When the label of feature mapping uses the letter \texttt{z}, it means that it only uses rotations around the z-axis. The corresponding operator is defined in \cref{eq:PauliZ}, and an example of such a mapping is displayed in \cref{fig:z}. When we talk about rotations, we mean them with respect to the visualization of the single qubit state on the Bloch sphere, which is shown in \cref{fig:bloch}. When there is a \texttt{zz} in a label, it means that an operator described in \Cref{eq:zz} is used; the example for this case is shown in \cref{fig:zz}. Furthermore, when a keyword \texttt{pauli} is used in the label, that means that more than one operator is used and these sets consist of operators described in \cref{eq:PauliX,eq:PauliY,eq:PauliZ} and shown in \cref{fig:p}. 

When suffix \texttt{linear} is used, it means that linear entanglement has been set to entangle the qubits, same as when \texttt{full}, which means full entanglement scheme. The number of repetitions is defined by the number after the keyword \texttt{reps}.

The feature mappings were chosen in a systematic way to test different levels of expressivity, from simple and unentangled baselines to more expressive multi-axis and entangled constructions. This structured selection enables a controlled comparison of how increasing encoding complexity and geometric richness affect learning dynamics and classification performance within an otherwise fixed hybrid architecture.

Out of nine tested mappings, only the \texttt{pauli\_xyz\_1\_rep} configuration resulted in high training and validation accuracy, while the remaining variants significantly hindered the model’s learning as seen in \cref{tab:accuracies}, inappropriate feature maps led to underfitting and poor generalization, as the accuracies of the model remained between $28\%$ and $36\%$, even when post-quantum-layer silhouette scores were moderately high numbers. This highlights that good separability alone does not guarantee strong model performance—accurate alignment between the feature space and the task-specific structure is essential.

\begin{table}[ht!]
\centering
\caption{Training and Validation Accuracies for Different Feature Maps}
\label{tab:accuracies}
\begin{tabular}{lcc}
\toprule
\textbf{Feature Map} & \textbf{Training Accuracy} & \textbf{Validation Accuracy} \\
\midrule
\texttt{zz\_reps\_2\_linear} & 0.2832 & 0.3340 \\
\texttt{z\_reps\_2} & 0.3193 & 0.3291 \\
\texttt{pauli\_z\_yy\_zxz\_linear} & 0.3145 & 0.3320 \\
\texttt{pauli\_xyz\_1\_rep} & \textbf{0.9082} & \textbf{0.9014} \\
\texttt{zz\_reps\_3\_full} & 0.3350 & 0.3235 \\
\texttt{zz\_reps\_1\_linear} & 0.3262 & 0.3174 \\
\texttt{pauli\_z\_yy\_zxz\_rep\_2} & 0.3271 & 0.3301 \\
\texttt{z\_reps\_1} & 0.3301 & 0.3408 \\
\texttt{z\_reps\_3} & 0.3125 & 0.3535 \\
\bottomrule
\end{tabular}
\end{table}

\begin{figure}[htbp]
    \centering

    \begin{subfigure}[t]{0.49\textwidth}
        \centering
        \includegraphics[width=\textwidth]{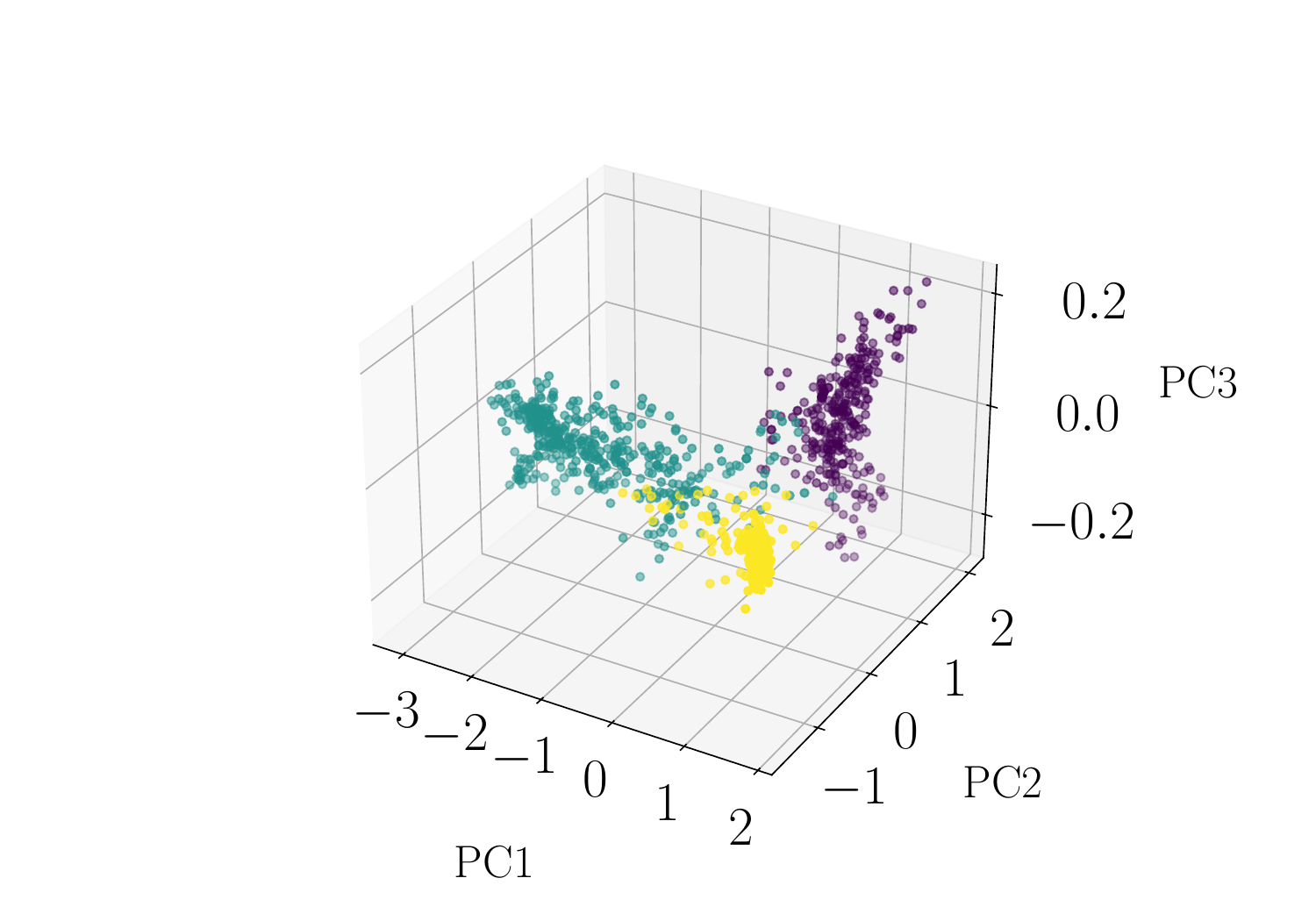}
        \caption{Principal Component Analysis After Classical Part, Color-coded by Fitted Values}
    \end{subfigure}
    \hfill
    \begin{subfigure}[t]{0.49\textwidth}
        \centering
        \includegraphics[width=\textwidth]{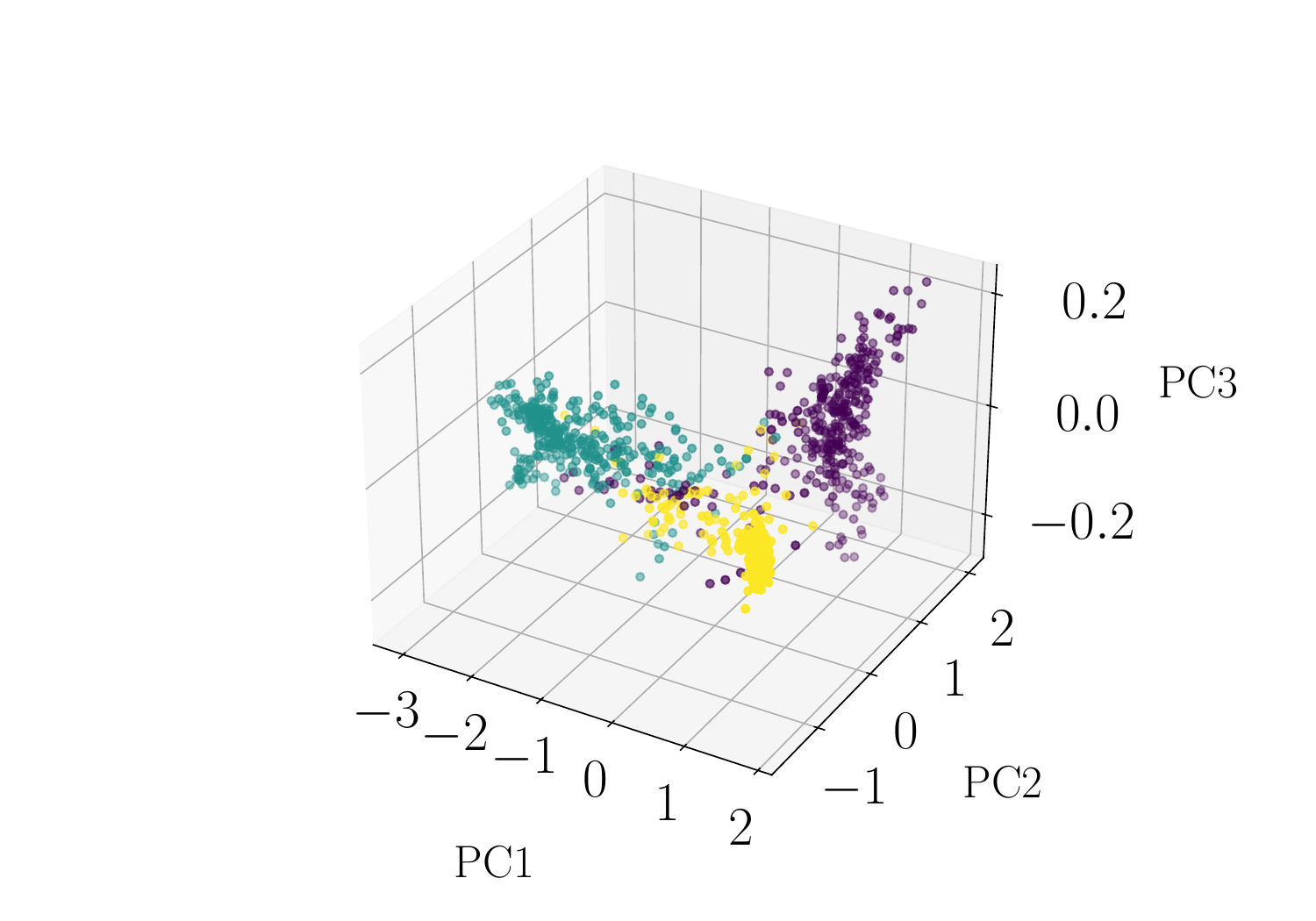}
        \caption{Principal Component Analysis After Classical Part, Color-coded by Training Values}
    \end{subfigure}

    \vspace{1em}

    \begin{subfigure}[t]{0.45\textwidth}
        \centering
        \includegraphics[width=\textwidth]{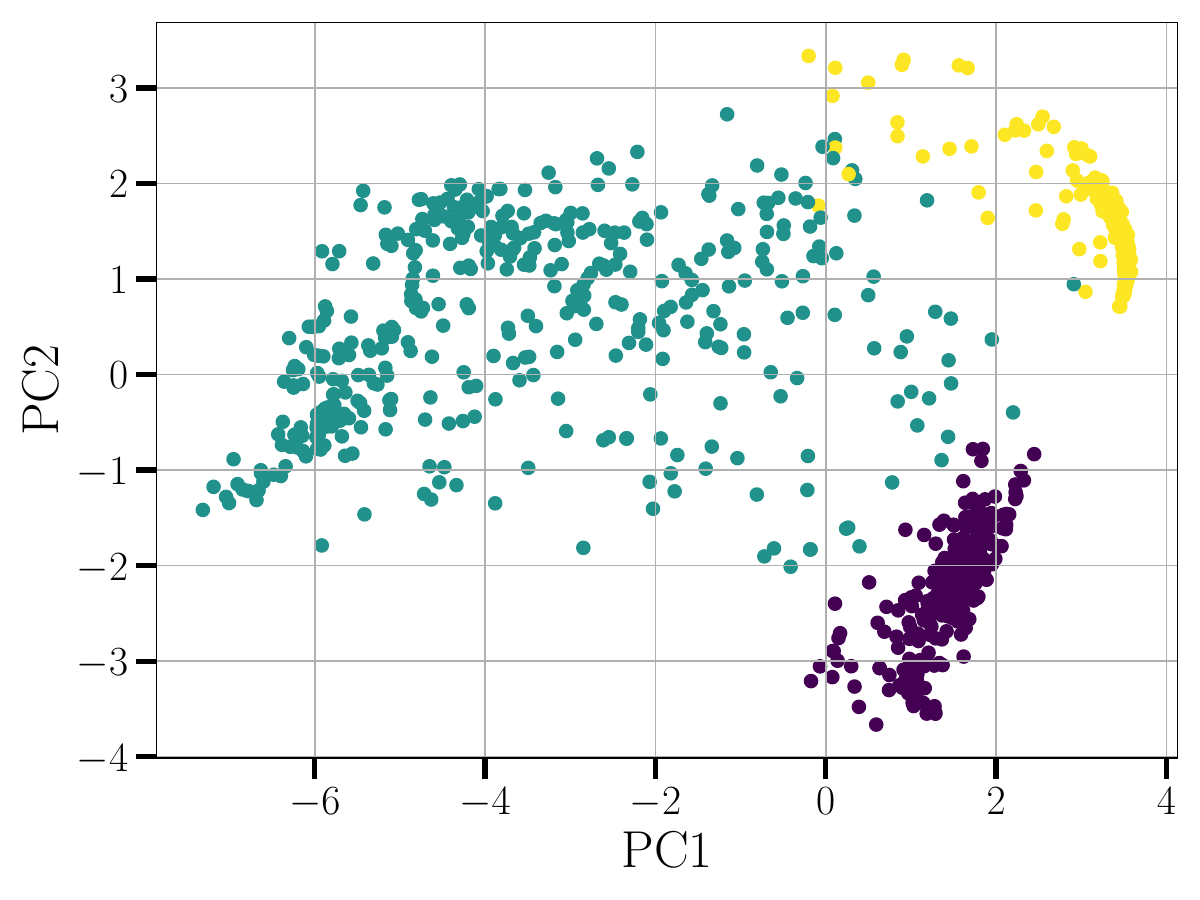}
        \caption{Principal Component Analysis After Feature Mapping, Color-coded by Fitted Values}
    \end{subfigure}
    \hfill
    \begin{subfigure}[t]{0.45\textwidth}
        \centering
        \includegraphics[width=\textwidth]{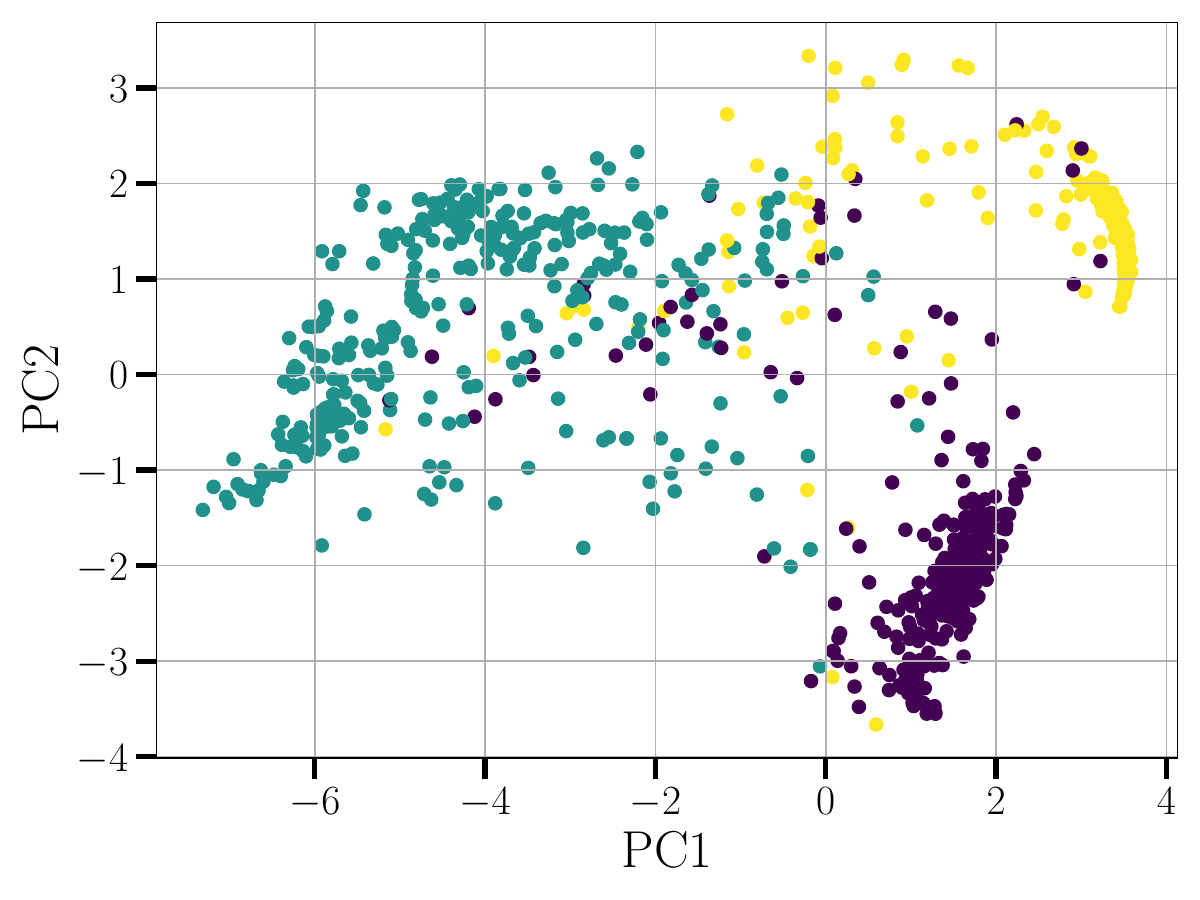}
        \caption{Principal Component Analysis After Feature Mapping, Color-coded by Training Values}
    \end{subfigure}

    \vspace{1em}

    \begin{subfigure}[t]{0.45\textwidth}
        \centering
        \includegraphics[width=\textwidth]{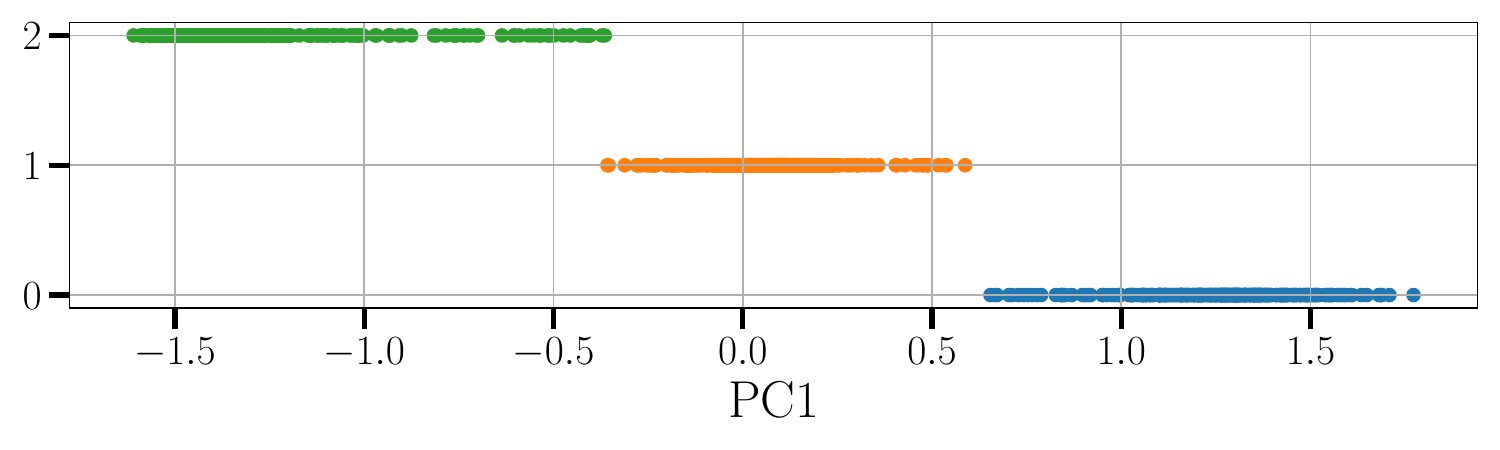}
        \caption{Principal Component Analysis After QNN, Color-coded by Fitted Values}
    \end{subfigure}
    \hfill
    \begin{subfigure}[t]{0.45\textwidth}
        \centering
        \includegraphics[width=\textwidth]{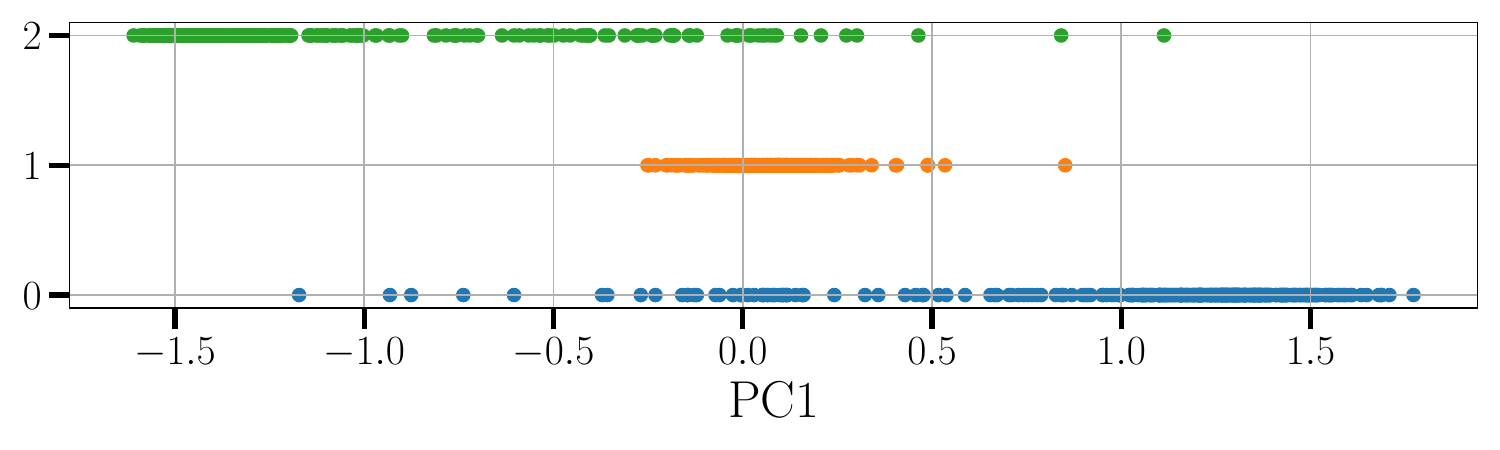}
        \caption{Principal Component Analysis After QNN, Color-coded by Training Values}
    \end{subfigure}

    \caption{PCA Progression for \texttt{pauli\_xyz\_1\_rep}}
    \label{fig:pauli_xyz_1_rep_train}
\end{figure}

\begin{figure}[htbp]
    \centering

    \begin{subfigure}[t]{0.49\textwidth}
        \centering
        \includegraphics[width=\textwidth]{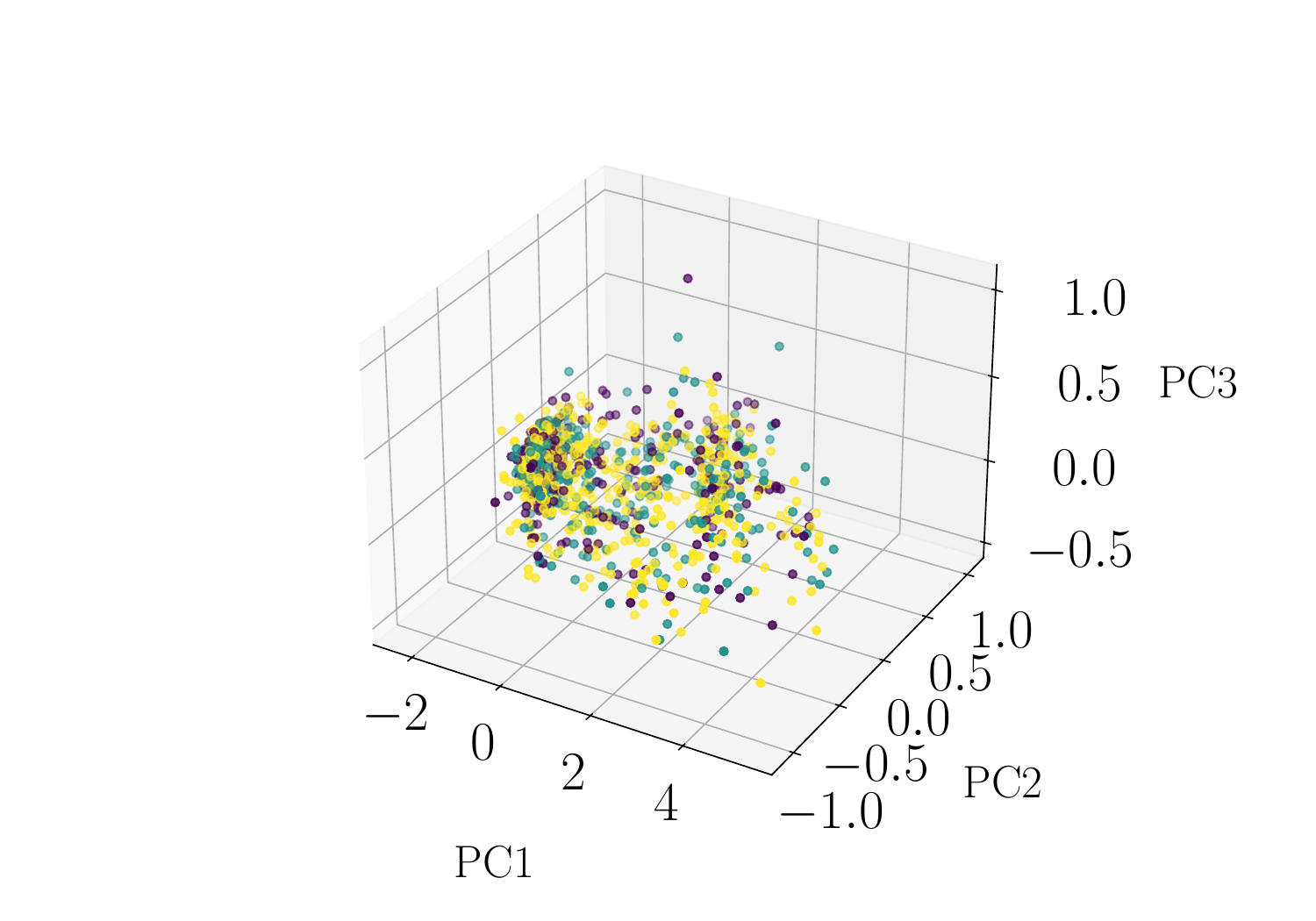}
        \caption{Principal Component Analysis After Classical Part, Color-coded by Fitted Values}
    \end{subfigure}
    \hfill
    \begin{subfigure}[t]{0.49\textwidth}
        \centering
        \includegraphics[width=\textwidth]{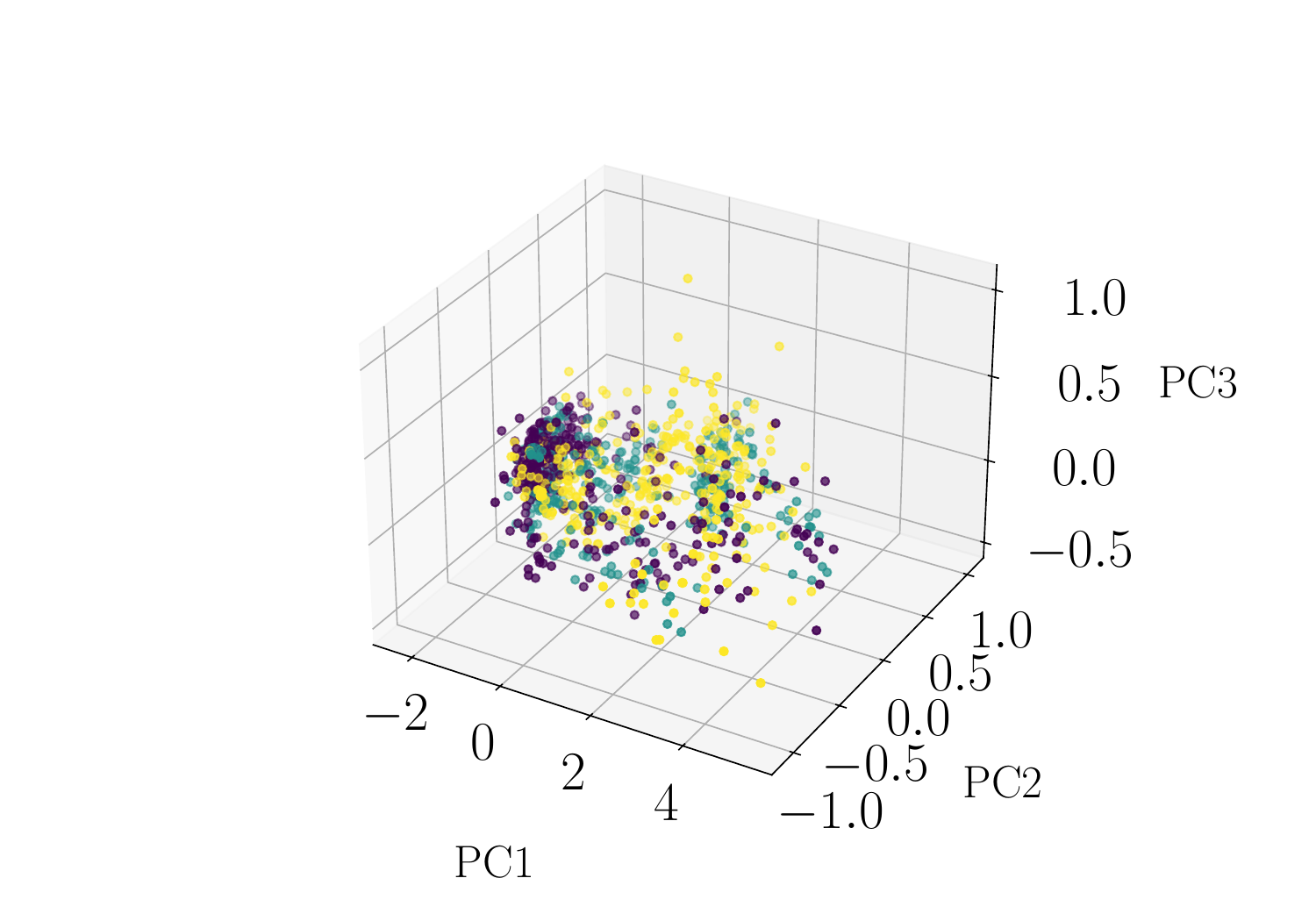}
        \caption{Principal Component Analysis After Classical Part, Color-coded by Training Values}
    \end{subfigure}

    \vspace{1em}

    \begin{subfigure}[t]{0.45\textwidth}
        \centering
        \includegraphics[width=\textwidth]{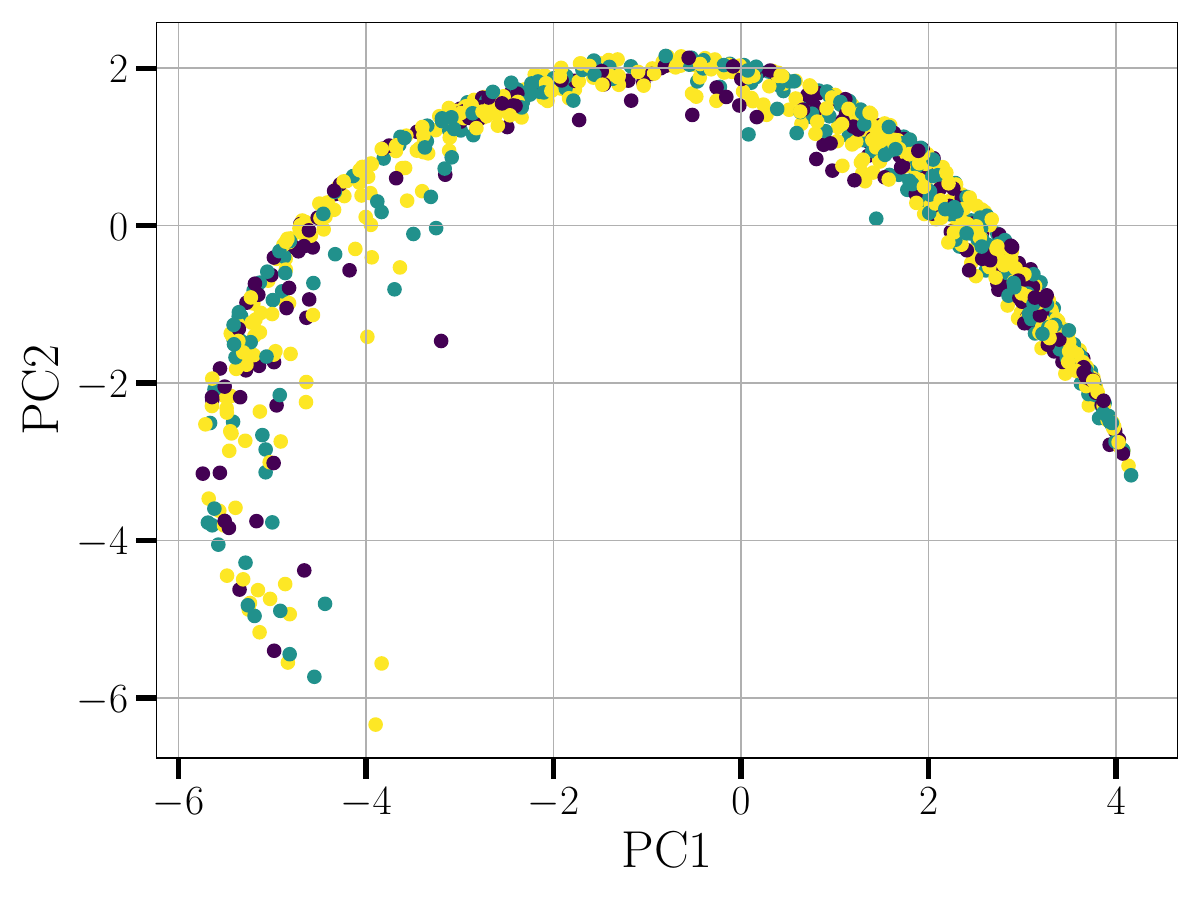}
        \caption{Principal Component Analysis After Feature Mapping, Color-coded by Fitted Values}
    \end{subfigure}
    \hfill
    \begin{subfigure}[t]{0.45\textwidth}
        \centering
        \includegraphics[width=\textwidth]{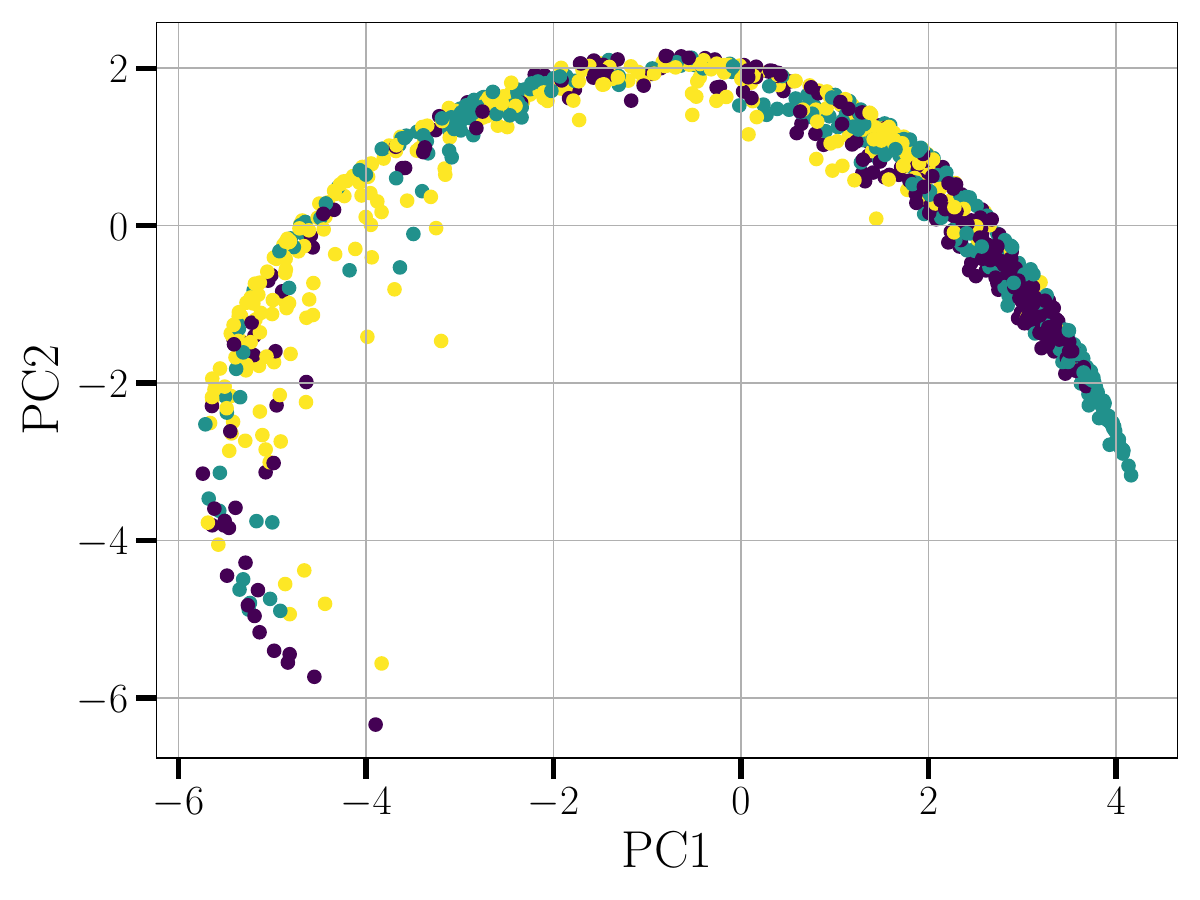}
        \caption{Principal Component Analysis After Feature Mapping, Color-coded by Training Values}
    \end{subfigure}

    \vspace{1em}

    \begin{subfigure}[t]{0.45\textwidth}
        \centering
        \includegraphics[width=\textwidth]{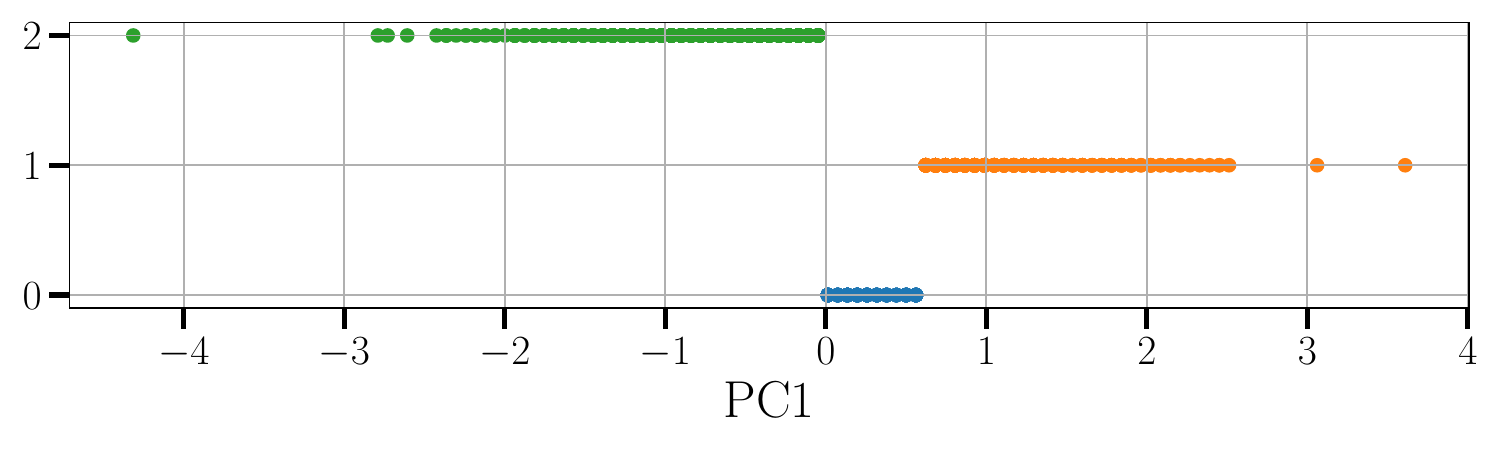}
        \caption{Principal Component Analysis After QNN, Color-coded by Fitted Values}
    \end{subfigure}
    \hfill
    \begin{subfigure}[t]{0.45\textwidth}
        \centering
        \includegraphics[width=\textwidth]{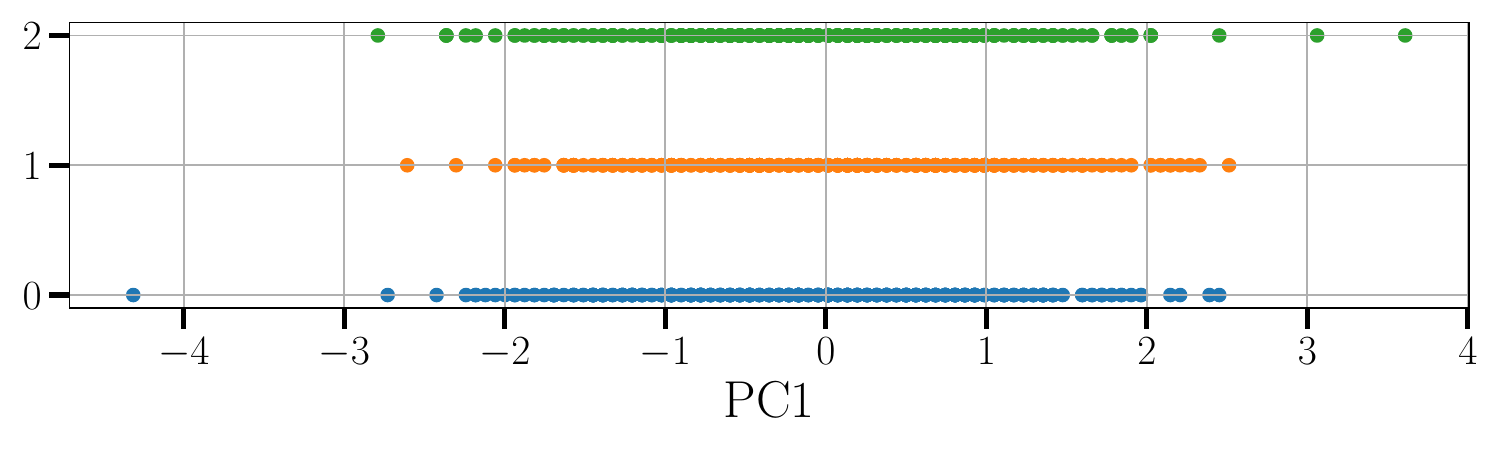}
        \caption{Principal Component Analysis After QNN, Color-coded by Training Values}
    \end{subfigure}

    \caption{PCA Progression for \texttt{z\_feature\_map\_reps\_1}}
    \label{fig:z_feature_map_reps_1_train}
\end{figure}

\begin{figure}[htbp]
    \centering

    \begin{subfigure}[t]{0.49\textwidth}
        \centering
        \includegraphics[width=\textwidth]{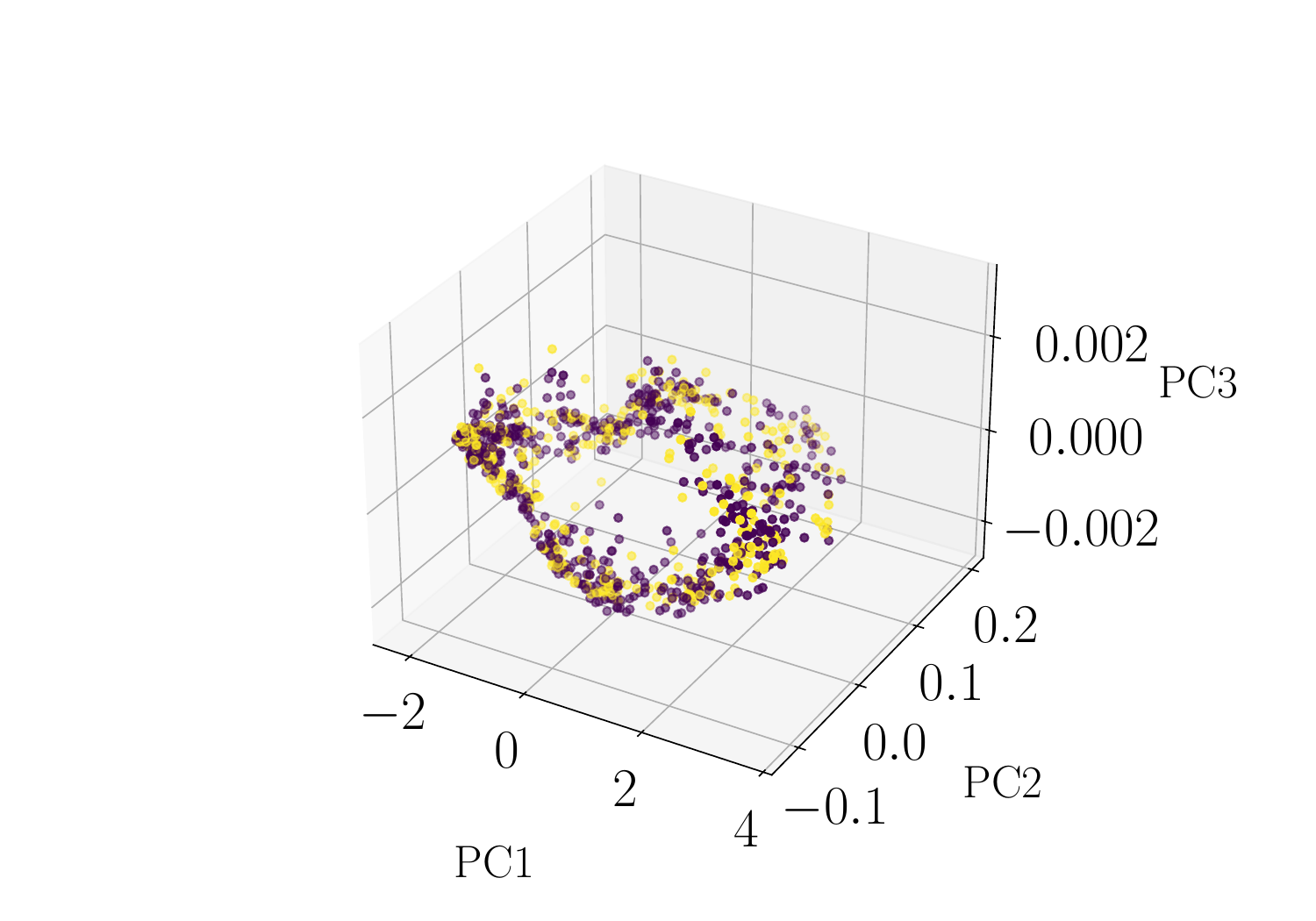}
        \caption{Principal Component Analysis After Classical Part, Color-coded by Fitted Values}
    \end{subfigure}
    \hfill
    \begin{subfigure}[t]{0.49\textwidth}
        \centering
        \includegraphics[width=\textwidth]{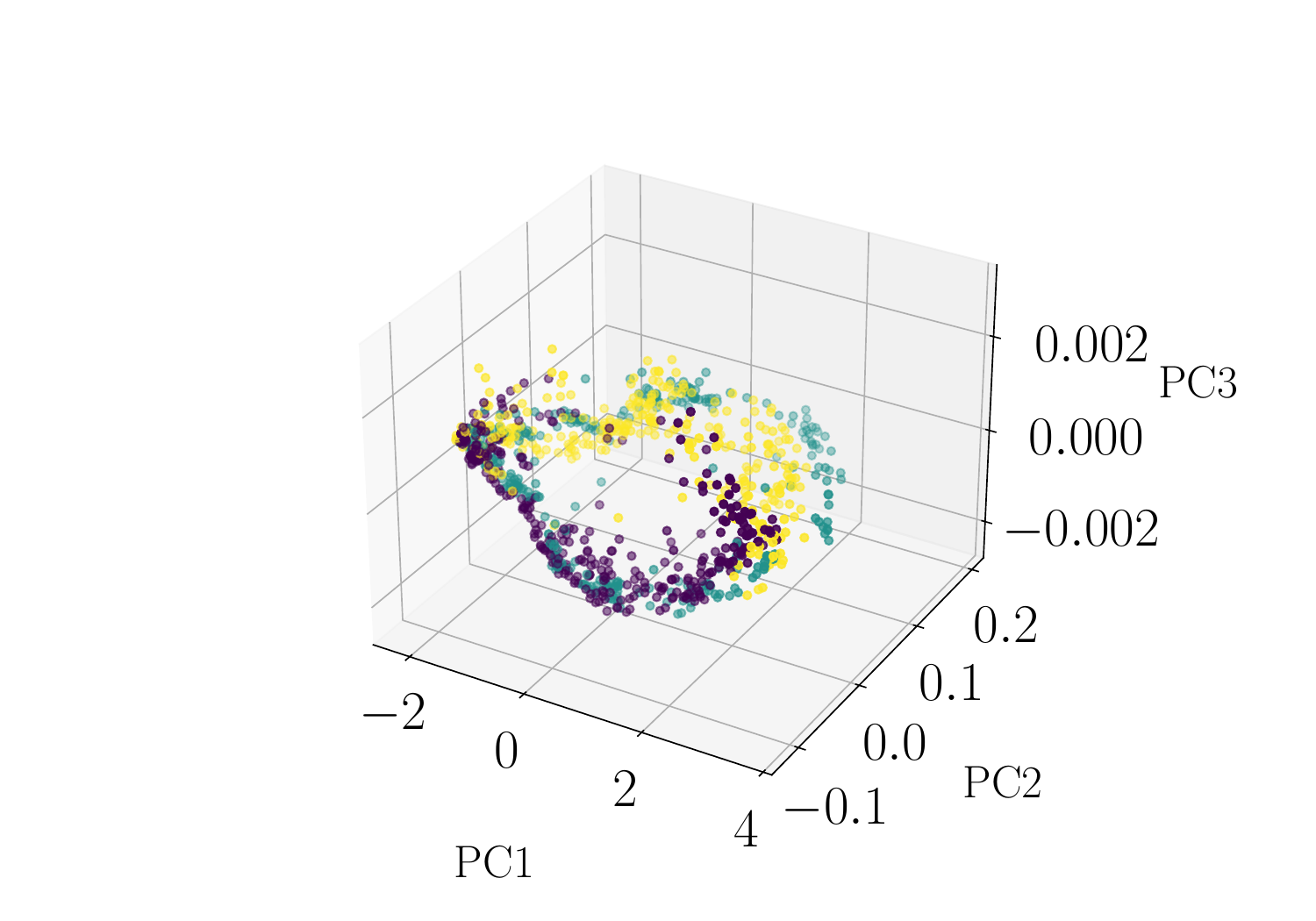}
        \caption{Principal Component Analysis After Classical Part, Color-coded by Training Values}
    \end{subfigure}

    \vspace{1em}

    \begin{subfigure}[t]{0.45\textwidth}
        \centering
        \includegraphics[width=\textwidth]{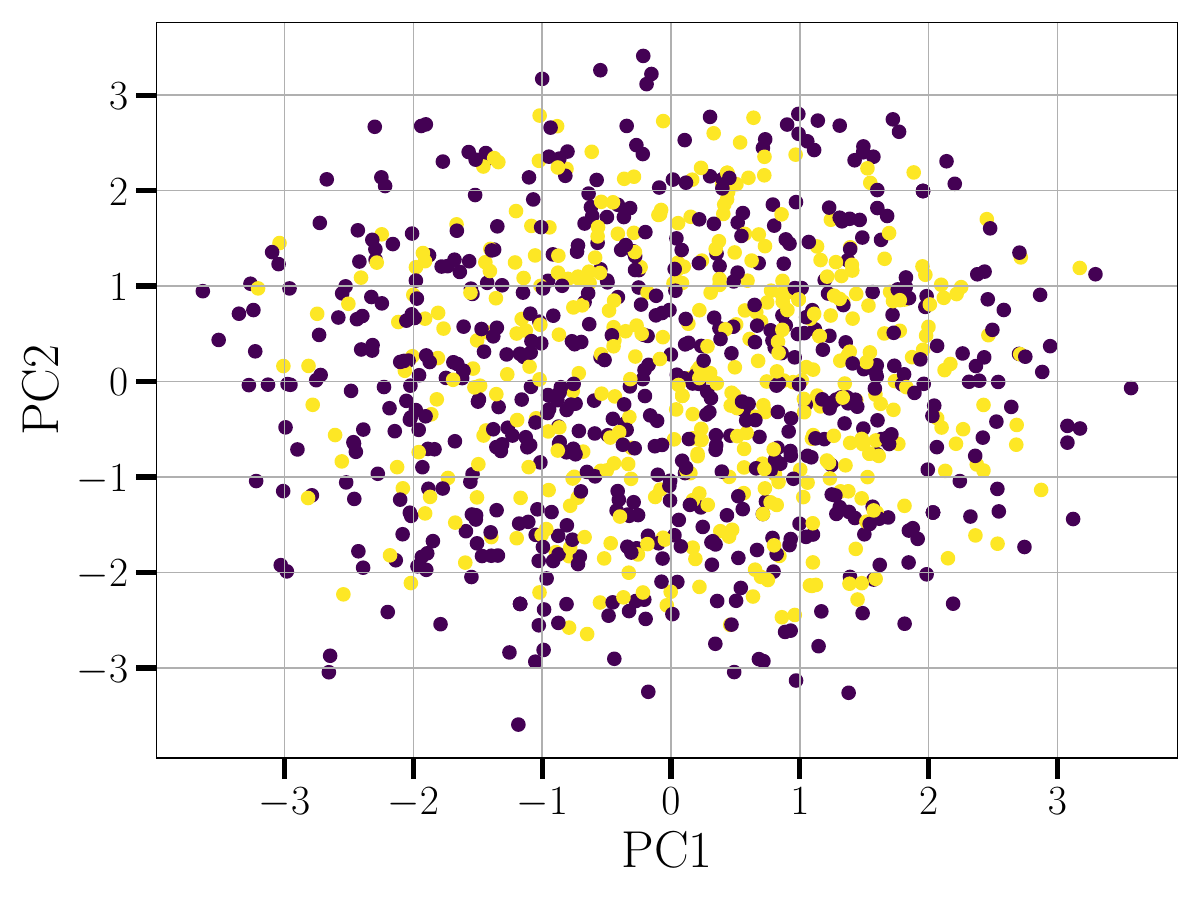}
        \caption{Principal Component Analysis After Feature Mapping, Color-coded by Fitted Values}
    \end{subfigure}
    \hfill
    \begin{subfigure}[t]{0.45\textwidth}
        \centering
        \includegraphics[width=\textwidth]{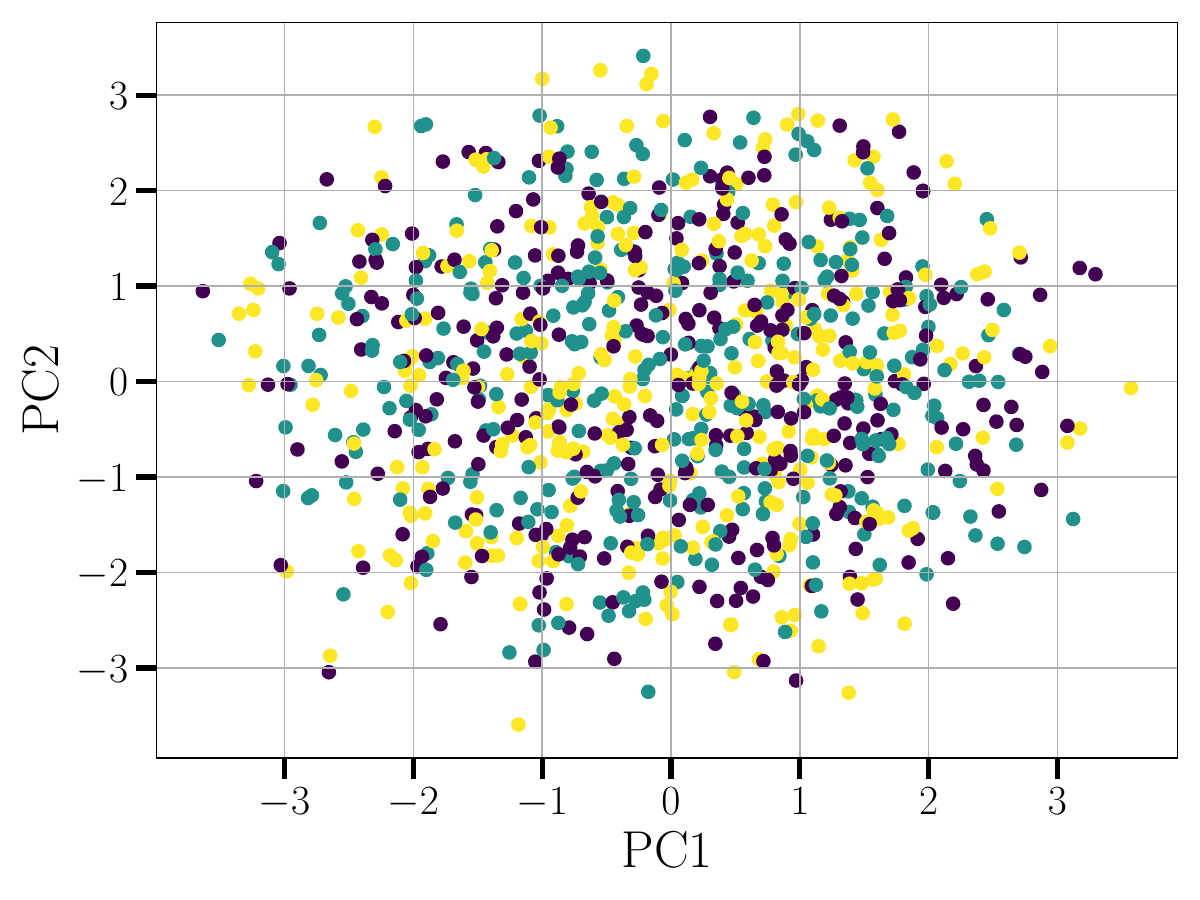}
        \caption{Principal Component Analysis After Feature Mapping, Color-coded by Training Values}
    \end{subfigure}

    \vspace{1em}

    \begin{subfigure}[t]{0.45\textwidth}
        \centering
        \includegraphics[width=\textwidth]{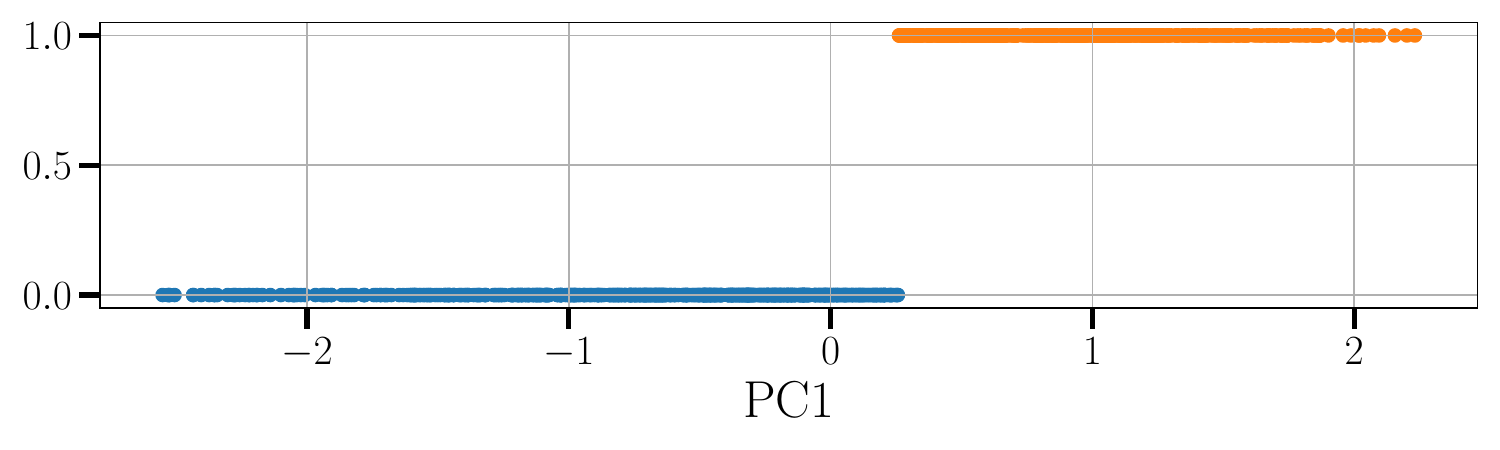}
        \caption{Principal Component Analysis After QNN, Color-coded by Fitted Values}
    \end{subfigure}
    \hfill
    \begin{subfigure}[t]{0.45\textwidth}
        \centering
        \includegraphics[width=\textwidth]{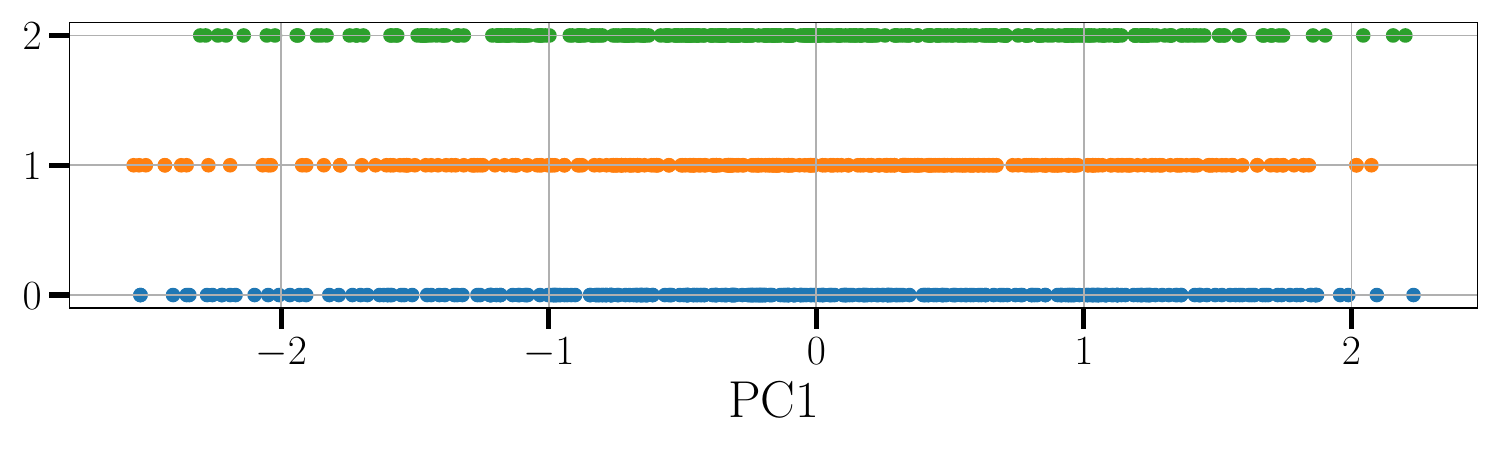}
        \caption{Principal Component Analysis After QNN, Color-coded by Training Values}
    \end{subfigure}

    \caption{PCA Progression for \texttt{zz\_feature\_map\_reps\_3\_full}}
    \label{fig:zz_feature_map_reps_3_full_train}
\end{figure}

While the accuracies around $30\%$ are consistent with near-random guessing for a three-class classification task, the same training pipeline has been used for all feature maps and successfully converged for \texttt{pauli\_xyz\_1\_rep}, confirming the correctness of the implementation. Thus, the poor performance of most mappings is attributed to limited expressivity, as when feature maps are dominated by $Z$ and $ZZ$ operators, the state rotates along only one axis when visualized on the Bloch sphere. This means that the different classes remain weakly distinguishable by the variational classifier. In contrast, \texttt{pauli\_xyz\_1\_rep} employs non-commuting Pauli operators across multiple axes, producing a richer embedding that enables effective class separation. This highlights that geometric separability, that is visualized in \cref{fig:pauli_xyz_1_rep_train,fig:z_feature_map_reps_1_train,fig:zz_feature_map_reps_3_full_train}, alone is insufficient and feature mappings must be selected with care.

To further analyze the behavior of different mappings, we performed PCA at three stages of the model: after the dimensionality reduction, after the feature mapping, and after the quantum neural network. For the successful \texttt{pauli\_xyz\_1\_rep}, which is shown in \cref{fig:pauli_xyz_1_rep_train}, the data began to show class-wise separation already after the classical layer, which was then amplified by the feature map and the quantum circuit. The silhouette score reached $0.80$ post-QNN, \cref{tab:pca_qnn_train}, indicating strong cluster separability aligned with the classification task. In contrast, simpler mappings like \texttt{z\_reps\_1} resulted in ring-like \gls{pca} structures as illustrated in \cref{fig:z_feature_map_reps_1_train}, a consequence of Z-rotations on the Bloch sphere. These transformations preserved the Z-coordinate of the qubit states, preventing the model from achieving meaningful data separation in later stages.

More complex encodings, such as \texttt{zz\_reps\_3\_full}, were also explored in an attempt to improve performance, as shown in \cref{fig:zz_feature_map_reps_3_full_train}. However, this approach led to dimensionality collapse, meaning that the model failed to distinguish between all three outputs and sorted the data into only two categories, with the model failing to distinguish between more than two output classes. Although the silhouette score post-QNN appeared to be high numbers, \cref{tab:pca_qnn_train}, a deeper inspection revealed that the learned features were misaligned with the actual training labels.

This misalignment becomes evident when comparing the silhouette scores for the fitted outputs and the original training data. While the post-QNN fitted representations yielded a silhouette score of $0.57$, suggesting seemingly well-separated clusters, the silhouette score on the actual training inputs dropped to $-0.02$, indicating that the learned representations did not align with the true label structure. In other words, the model produced confident-looking outputs that did not reflect meaningful class separability in the input space. This is further corroborated by the low training and validation accuracies ($0.3350$ and $0.3235$, respectively), revealing that the model failed to learn a mapping from input features to labels. The results suggest that the increased depth and entanglement in \texttt{zz\_reps\_3\_full} may have caused the QNN to overfit to latent patterns that are not relevant for the supervised task, ultimately leading to dimensionality collapse and degraded classification performance.

This observation emphasizes the need for principled feature map design. Neither simplistic nor overly complex mappings consistently yield good results—rather, effective feature mappings must strike a balance between expressivity and compatibility with the underlying data structure and model architecture, where expressivity is the mapping's ability to distinctively and reasonably encode classical inputs into a quantum circuit.

\cref{tab:pca_class_train} summarizes the \gls{pca} results and silhouette scores obtained after the classical (pre-quantum) part of the hybrid model. In nearly all cases, the first two or three principal components capture over 95\% of the variance, demonstrating that the classical layers effectively compress the input data into a low-dimensional subspace. However, this dimensionality reduction does not translate into class separability, as indicated by low or even negative silhouette scores, from $-0.07$ to $0.32$ for most feature maps, as can be seen in \cref{tab:pca_class_train}. The one clear exception is \texttt{pauli\_xyz\_1\_rep}, which achieves silhouette scores of 0.70 (fitted) and 0.62 (training), suggesting that this configuration already facilitates some degree of class-wise clustering at the classical stage. These results reveal that while classical processing reduces data complexity, it is insufficient on its own for effective class separation, hinting at the need for carefully designed quantum layers to enhance discriminative structure, i.e., a structure where there is a clear distinction between classes.

\cref{tab:pca_feat_train} presents the \gls{pca} variances and silhouette scores for the training data immediately after the feature mapping stage. A general trend can be observed, where the proportion of variance captured by the first two principal components decreased compared to the classical stage, indicating increased data complexity introduced by the non-linear feature encodings. However, the silhouette scores remained low or even negative for most mappings, from $-0.04$ to $0.08$, implying limited class separability at this point. The only clear exception was the \texttt{pauli\_xyz\_1\_rep}, which yielded relatively high silhouette scores of $0.55$ (fitted) and $0.48$ (training), suggesting the emergence of some degree of meaningful cluster structure. Despite this, even moderately successful feature mappings alone were insufficient for full separability, confirming that the subsequent trainable quantum layer plays a crucial role in refining the representation space.

\cref{tab:pca_qnn_train} presents the \gls{pca} variance and silhouette scores after the quantum layer has been applied. Notably, in all cases, the variance collapsed into a single dominant component, noted by \gls{pca} variance of $1.0$, indicating that the quantum circuit significantly restructured the feature space into a highly compressed representation. This is to be expected as the output of the quantum layer is a single scalar number. This suggests that the QNN effectively projects the data into a lower-dimensional but potentially more expressive space, as the transformation through feature mapping distributes the classes across the Hilbert space. More importantly, for several feature maps, the silhouette scores improved significantly compared to earlier stages. For instance, \texttt{pauli\_xyz\_1\_rep} reached a silhouette score of $0.80 (fitted)$, indicating well-separated clusters aligned with the model’s decision boundaries, meaning that the outputs for different classes are inside different, independent intervals. Other mappings, such as \texttt{z\_feature\_map\_reps\_3} and \texttt{z\_feature\_map\_reps\_2}, also showed high silhouette scores ($0.75$ and $0.65$ respectively). However, a consistent pattern is observed as silhouette scores for fitted values were often high numbers, while those for training labels remained low or negative numbers. This highlights a potential overfitting issue or a misalignment between learned clustering and true class labels, so either the model lost its generalization ability, or divided data into separate clusters, that are artifital with no bearing to the classification provided by training labels/ This emphasizes the necessity of selecting feature maps that not only enable the QNN to find separable clusters but also align them with the actual classification task, as we want the clusters to correspond to the real classification provided a priori.

\section{Conclusions}\label{sec:conclusions}
This work has demonstrated that the performance of hybrid quantum-classical neural networks is deeply influenced by the design of the quantum layer, particularly in terms of ansatz depth and feature mapping. Through systematic experiments, we identified several key patterns that can guide future applications of quantum machine learning in classification tasks.

First, increasing the depth of the quantum ansatz improves model performance in multiple ways. Models with deeper ansatz exhibited higher validation accuracy, better generalization, and smoother training dynamics. Notably, although deeper models showed slightly slower initial learning, this was quickly compensated for by improved convergence and robustness. The most expressive model tested—using three ansatz repetitions—was the first to surpass 90\% accuracy and demonstrated the lowest fluctuation in both training and validation performance. Interestingly, these deeper circuits also acted as implicit regularizers, reducing overfitting without requiring explicit penalties. This was reflected in a stability ratio below one, indicating greater stability on validation data than on training data.

However, the improvements gained from increasing the number of ansatz block repetitions were found to be diminishing. The transition from one to two repetitions led to the most substantial improvement, while adding a third showed more modest gains. This suggests that beyond a certain point, increasing circuit complexity may yield limited or even adverse effects, at least for the model of this structure. Indeed, unnecessary depth can exacerbate optimization difficulties due to an increasingly complex loss landscape, particularly under limited training budgets.

The choice of feature map proved to be even more critical. Of the nine feature maps tested, only one \texttt{pauli\_xyz\_1\_rep} enabled the model to learn effectively. This feature map employed rotations around multiple Pauli axes, which appeared essential for capturing rich data structure. In contrast, feature maps based solely on Z rotations, even when repeated or entangled, consistently underperformed and in some cases failed to distinguish between all classes. Moreover, adding complexity without careful design—such as using full entanglement or high repetition counts—often degrades performance due to dimensional collapse or poor trainability. These findings emphasize that a thoughtful, task-aligned design of feature maps is crucial in hybrid settings.

Additionally, our analysis using \gls{pca} and silhouette scores showed that successful models transformed the data into well-separated and structured feature spaces after the quantum layer, whereas underperforming models produced chaotic or overly compressed distributions. 

Lastly, it is essential to allow sufficient training time when evaluating hybrid models. Deeper quantum layers may exhibit slower initial progress, and premature judgments may lead to the rejection of well-structured, high-performing models. Overall, the findings underscore that the architecture and design of the quantum component not only influence final accuracy but also shape the learning dynamics throughout training. To achieve robust performance in hybrid quantum-classical classifiers, we recommend the use of multi-axis feature maps, moderate circuit depth, and careful monitoring of clustering and generalization behavior throughout training.

While the presented study offers valuable insights into the effects of ansatz depth and feature mapping on hybrid quantum-classical models, it is constrained by several limitations. Most notably, the investigation is limited to a specific hybrid architecture—convolutional layers followed by a quantum layer with the \texttt{TwoLocal} ansatz—and evaluated on a single, domain-specific dataset derived from bivariate causal inference. As a result, the generalizability of the observed trends to other types of hybrid models or tasks remains unverified. Moreover, the study explores only a fixed range of circuit depths and feature maps. Although this allows for controlled comparisons, it leaves open the question of whether other, potentially more effective architectures exist beyond the tested configurations. Additionally, the evaluation framework, while comprehensive, primarily relies on \gls{pca} and silhouette scores, which may not fully capture the quantum circuit’s representational complexity or expressivity in high-dimensional Hilbert space.

Future research can address these limitations by expanding the architectural scope of hybrid models and introducing alternative quantum circuit designs, such as hardware-efficient ansatz tailored to specific quantum hardware topologies or domain-inspired encodings. Incorporating tasks with different data modalities and evaluating performance across multiple benchmarks will help assess the generalizability of the findings. Additionally, automated architecture search methods, such as quantum neural architecture search~\cite{duong2022quantum}, may be employed to explore a broader design space of quantum layers. From an evaluation perspective, supplementing \gls{pca}-based separability metrics with entanglement entropy~\cite{calabrese2004entanglement}, Fisher information~\cite{rissanen2002fisher}, or kernel-based expressivity measures~\cite{jager2023universal} could provide deeper insights into quantum model behavior. Finally, integrating noise modeling and hardware simulation would bridge the gap between idealized training conditions and realistic performance on near-term quantum devices.

Future work will build on the current findings by exploring a broader range of quantum circuit architectures and applying the hybrid model to more complex datasets to evaluate scalability and robustness. These efforts will aim to deepen our understanding of how different quantum components interact with classical layers under varying learning conditions. Also, we will extend the tested pipeline to include cross-validation and randomized train and test set splits, as well as more formal statistical tests to assess the significance of
performance differences. This will enable a more rigorous assessment of statistical robustness while directly extending from this work. An extended analysis featuring further metrics~\cite{illesova2025qmetric} will also be performed. The insights gained from these continued experiments will form the basis for future studies, where the extended analysis and results will be presented in greater depth to contribute to the growing body of research in quantum machine learning.

\section{Acknowledgments}
This work was supported by the Ministry of Education, Youth and Sports of the Czech Republic through the e-INFRA CZ (ID:90254).
Martin Beseda was supported by Italian Government (Ministero dell’Università e della Ricerca, PRIN 2022 PNRR)---cod.
P2022SELA7: ``RECHARGE: monitoRing, tEsting, and CHaracterization of performAnce Regressions''---D.D. n. 1205 del 28/7/2023. PG and TR gratefully acknowledge the funding support by program ``Excellence Initiative — Research University'' for the AGH University of Kraków, as well as the ARTIQ project ARTIQ/0004/2021.

\bibliographystyle{plainnat}
\bibliography{interactnlmsample}
\newpage
\appendix
\section{Tables}

\begin{table}[ht]
\centering
\caption{PCA Variance and Silhouette Scores After Feature Mapping on Training Data}
\label{tab:pca_feat_train}
\begin{tabular}{p{7cm} p{2cm} p{1cm}}
\toprule
\textbf{Feature Map} & \textbf{PC Variance [1, 2]} & \textbf{Silhouette Score} \\
\midrule
ZZ Feature Map Reps 2 Linear (Fitted) & [0.18, 0.16] & 0.04 \\
ZZ Feature Map Reps 2 Linear (Train) & [0.18, 0.16] & -0.01 \\
Z Feature Map Reps 2 (Fitted) & [0.23, 0.20] & 0.08 \\
Z Feature Map Reps 2 (Train) & [0.23, 0.20] & 0.01 \\
Pauli Z YY ZXZ Linear (Fitted) & [0.08, 0.07] & 0.00 \\
Pauli Z YY ZXZ Linear (Train) & [0.08, 0.07] & -0.00 \\
Pauli XYZ 1 Rep (Fitted) & [0.68, 0.18] & 0.55 \\
Pauli XYZ 1 Rep (Train) & [0.68, 0.18] & 0.48 \\
ZZ Feature Map Reps 3 Full (Fitted) & [0.12, 0.11] & 0.00 \\
ZZ Feature Map Reps 3 Full (Train) & [0.12, 0.11] & -0.00 \\
ZZ Feature Map Reps 1 No Entanglement (Fitted) & [0.68, 0.26] & -0.02 \\
ZZ Feature Map Reps 1 No Entanglement (Train) & [0.68, 0.26] & -0.02 \\
Pauli Z YY ZXZ Rep 2 (Fitted) & [0.08, 0.07] & -0.00 \\
Pauli Z YY ZXZ Rep 2 (Train) & [0.08, 0.07] & -0.00 \\
Z Feature Map Reps 1 (Fitted) & [0.69, 0.18] & -0.01 \\
Z Feature Map Reps 1 (Train) & [0.69, 0.18] & -0.00 \\
Z Feature Map Reps 3 (Fitted) & [0.28, 0.25] & -0.02 \\
Z Feature Map Reps 3 (Train) & [0.28, 0.25] & -0.00 \\
\bottomrule
\end{tabular}
\end{table}

\begin{table}[ht]
\centering
\caption{PCA Variance and Silhouette Scores After Classical Part on Training Data}
\label{tab:pca_class_train}
\begin{tabular}{p{7cm} p{2cm} p{1cm}}
\toprule
\textbf{Feature Map} & \textbf{PC Variance [1, 2, 3]} & \textbf{Silhouette Score} \\
\midrule
ZZ Feature Map Reps 2 Linear (Fitted) & [0.84, 0.16, 0.01] & 0.20 \\
ZZ Feature Map Reps 2 Linear (Train) & [0.84, 0.16, 0.01] & 0.02 \\
Z Feature Map Reps 2 (Fitted) & [0.97, 0.02, 0.01] & 0.05 \\
Z Feature Map Reps 2 (Train) & [0.97, 0.02, 0.01] & -0.02 \\
Pauli Z YY ZXZ Linear (Fitted) & [1.00, 0.00, 0.00] & -0.00 \\
Pauli Z YY ZXZ Linear (Train) & [1.00, 0.00, 0.00] & -0.07 \\
Pauli XYZ 1 Rep (Fitted) & [0.67, 0.33, 0.00] & 0.70 \\
Pauli XYZ 1 Rep (Train) & [0.67, 0.33, 0.00] & 0.62 \\
ZZ Feature Map Reps 3 Full (Fitted) & [1.00, 0.00, 0.00] & 0.00 \\
ZZ Feature Map Reps 3 Full (Train) & [1.00, 0.00, 0.00] & -0.03 \\
ZZ Feature Map Reps 1 No Entanglement (Fitted) & [0.78, 0.15, 0.07] & -0.01 \\
ZZ Feature Map Reps 1 No Entanglement (Train) & [0.78, 0.15, 0.07] & -0.01 \\
Pauli Z YY ZXZ Rep 2 (Fitted) & [1.00, 0.00, 0.00] & -0.02 \\
Pauli Z YY ZXZ Rep 2 (Train) & [1.00, 0.00, 0.00] & -0.04 \\
Z Feature Map Reps 1 (Fitted) & [0.95, 0.04, 0.01] & -0.02 \\
Z Feature Map Reps 1 (Train) & [0.95, 0.04, 0.01] & -0.02 \\
Z Feature Map Reps 3 (Fitted) & [0.96, 0.03, 0.00] & 0.32 \\
Z Feature Map Reps 3 (Train) & [0.96, 0.03, 0.00] & 0.01 \\
\bottomrule
\end{tabular}
\end{table}

\begin{table}[ht]
\centering
\caption{PCA Variance and Silhouette Scores After QNN on Training Data}
\label{tab:pca_qnn_train}
\begin{tabular}{p{7cm} p{2cm} p{1cm}}
\toprule
\textbf{Feature Map} & \textbf{PC Variance} & \textbf{Silhouette Score} \\
\midrule
ZZ Feature Map Reps 2 Linear (Fitted) & [1.00] & 0.57 \\
ZZ Feature Map Reps 2 Linear (Train) & [1.00] & -0.04 \\
Z Feature Map Reps 2 (Fitted) & [1.00] & 0.65 \\
Z Feature Map Reps 2 (Train) & [1.00] & -0.05 \\
Pauli Z YY ZXZ Linear (Fitted) & [1.00] & 0.49 \\
Pauli Z YY ZXZ Linear (Train) & [1.00] & -0.04 \\
Pauli XYZ 1 Rep (Fitted) & [1.00] & 0.80 \\
Pauli XYZ 1 Rep (Train) & [1.00] & 0.62 \\
ZZ Feature Map Reps 3 Full (Fitted) & [1.00] & 0.57 \\
ZZ Feature Map Reps 3 Full (Train) & [1.00] & -0.02 \\
ZZ Feature Map Reps 1 No Entanglement (Fitted) & [1.00] & 0.38 \\
ZZ Feature Map Reps 1 No Entanglement (Train) & [1.00] & -0.01 \\
Pauli Z YY ZXZ Rep 2 (Fitted) & [1.00] & 0.27 \\
Pauli Z YY ZXZ Rep 2 (Train) & [1.00] & -0.02 \\
Z Feature Map Reps 1 (Fitted) & [1.00] & 0.42 \\
Z Feature Map Reps 1 (Train) & [1.00] & -0.02 \\
Z Feature Map Reps 3 (Fitted) & [1.00] & 0.75 \\
Z Feature Map Reps 3 (Train) & [1.00] & -0.11 \\
\bottomrule
\end{tabular}
\end{table}

\FloatBarrier
\section{Feature Mappings}
\begin{figure}[h]
    \centering
    \includegraphics[width=0.95\linewidth]{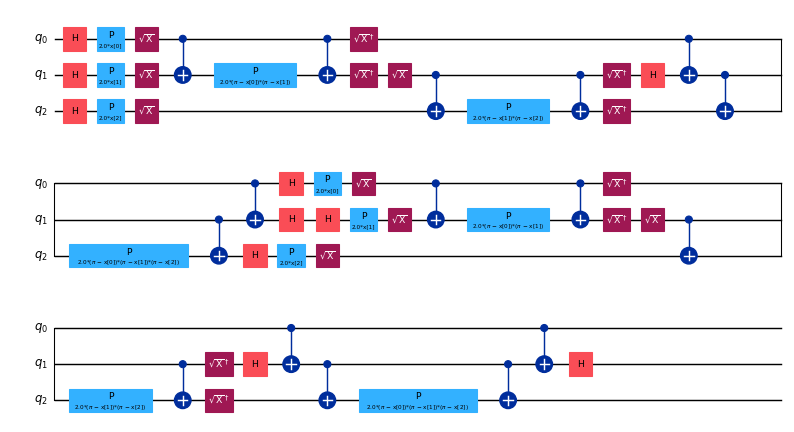}
    \caption{Feature mapping with different Pauli operators and linear entanglement}
    \label{fig:p}
\end{figure}
\begin{figure}[h]
    \centering
    \includegraphics[width=0.5\linewidth]{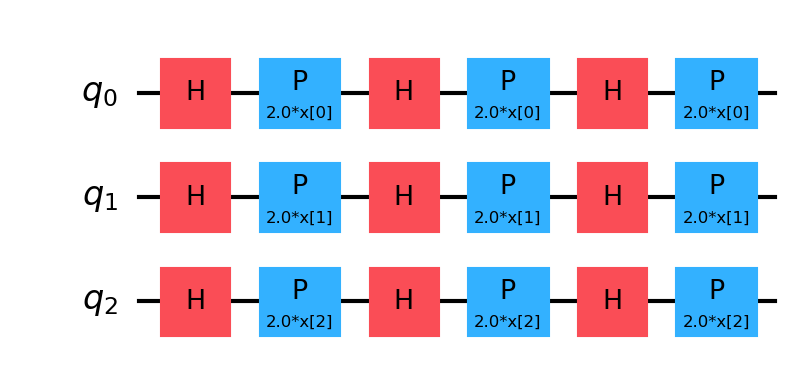}
    \caption{Feature mapping with Z rotations, 3 repetitions, and linear entanglement}
    \label{fig:z}
\end{figure}
\begin{figure}[h]
    \centering
    \includegraphics[width=0.7\linewidth]{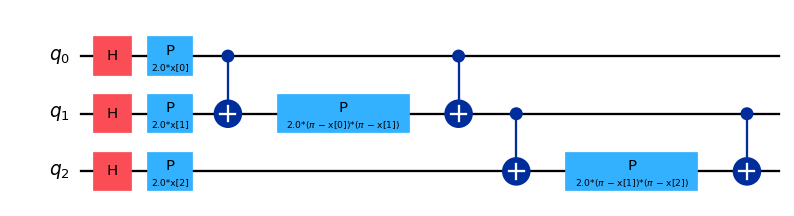}
    \caption{Feature mapping with ZZ operators and linear entanglement}
    \label{fig:zz}
\end{figure}

\end{document}